\documentclass[ showpacs,aps]{revtex4-1}
\usepackage{amsmath}
\usepackage{graphicx}
\usepackage{epstopdf}        

\begin{document}

\title{\bf Exact relativistic models of  conformastatic charged dust thick disks }


\author{Gonzalo Garc\'{\i}a-Reyes}
\email[e-mail: ]{ggarcia@utp.edu.co}
\affiliation{Departamento de F\'{\i}sica, Universidad Tecnol\'ogica de Pereira,
 A. A. 97, Pereira, Colombia}

 \begin{abstract}
We construct  relativistic models of  charged dust  thick disks  for  a particular conformastatic spacetime through a Miyamoto-Nagai transformation used in Newtonian gravity to model disk like  galaxies.   Two  simple families of thick disk models  and a family of thick annular disks  based on the field of 
an  extreme  Reissner-Nordstr\"om  black hole  and  a Morgan-Morgan-like metric   are considered.  The  electrogeodesic motion of test particles around the structures  are analyzed. Also the stability of the particles against radial perturbation is  studied using an extension of the Rayleigh criteria of stability of a fluid in rest in a
gravitational field. The models built satisfy all the energy conditions.    

\end{abstract}

\maketitle

\section{Introduction}

Axially symmetric exact solutions of Einstein field's equations  describing the field of a disk are important  in astrophysics as models of  certain stars, flat galaxies, accretion disks, the superposition of a black hole and a galaxy or an accretion
disk as in the case of quasars, and in general relativity as sources of exact solutions of Einstein equations.
Exact solutions  representing   static thin disks  were first studied by  Bonnor and Sackfield \cite{BS}
and Morgan and Morgan \cite{MM1, MM2}.   Several classes of exact solutions of the  Einstein  equations
corresponding to static thin disks have been
obtained by different authors
\cite{LP,CHGS,LO,LEM,BLK,BLP,GE}, while 
rotating  thin disks  were studied in \cite{BL,GL2}.
The exact superposition of a disk and a static black hole was first considered
by Lemos and Letelier  \cite{LL1,LL2}.  Static  thick disk models were considered by Gonz\'alez and Leterier  \cite{G-L-thick} and rotating thick disks by Voght and Letelier   \cite{V-L-thick}.

On the other hand, thin disks in presence
of electromagnetic field have been discussed as sources for 
Kerr-Newman fields  \cite{LBZ},
conformastationary metrics \cite{KBL},
and magnetostatic  axisymmetric fields  \cite{LET1,GG1,GG2,GG-CRIS,G-O}.  Now, have  been showed that 
gravitationally bound system such as stars, galaxies, and clusters of galaxies can be  positively charged \cite{Rosseland,Bally}, 
so that    electric fields can also be important in the study of self-gravitating system     not only  from the merely theoretical point of view but  also  astrophysical. Thus, in presence of a pure electric field, conformastatic  thin disks made of a  charged perfect fluid have been considered   in \cite{V-L-perfect},
and  composed of
electrically counterpoised dust \cite{majum,papa,bonnor98}, i.e. 
 with the electric charge density  equal to the mass density,  
in  reference  \cite{V-L-dust} for   nonaxisymmetric planar matter distribution,  
in \cite{G-A-P} for Morgan-Morgan-like  fields and in \cite{Lora}
for the superposition of  an   extreme  Reissner-Nordstr\"om  black hole and a annular disk. 

In this work we construct   exact relativistic models of thick disks made of a charged dust fluid for  a particular conformastatic spacetime through a Miyamoto-Nagai transformation used in Newtonian gravity to model disk like  galaxies \cite{Miyamoto,Nagai}, which adds  to the previous electrostatic  disks models
an additional degree of reality, its thickness.  Two  simple families of thick disk models  and a family of thick annular disks  based on the field of 
an  extreme  Reissner-Nordstr\"om  black hole  and  a Morgan-Morgan-like solution   are considered. The ring like matter distributions are generated   applying  a Kelvin transformation \cite{Kelvin1,Kelvin2}.   The structures  have no
boundary but as the energy density
decreases rapidly one can define a cutoff radius and, in principle, to
consider such structures as finite.

The paper is organized as follows.  In Sec. II we present
the formalism  to construct   models of  thick  disks  made of  charged dust for a   particular conformastic metric.
We also analysis the electrogeodesic motion of  charged  test particles   around  the disks and the stability of the orbits against radial perturbation using an extension of the Rayleigh criteria of stability of a fluid in rest in a
gravitational field.
In Secs. III-V  two simple family  of  charged dust  thick disks and a family of thick annular disks  are considered based    on the field of 
an  extreme  Reissner-Nordstr\"om  black hole  and  a Morgan-Morgan-like solution.  Finally, in Sec. VI we summarize and discuss the results obtained.


\section{Einstein-Maxwell equations  and disks } 

We consider a conformastatic spacetime in   cylindrical coordinates
 $(\varphi, R,z)$  and in the particular form
 \begin{equation}
ds^2 =  -(1-\psi)^{-2} dt^2 +  (1-\psi)^2  (R^2 d \varphi ^2 + d R^2 + dz^2  ). \label{eq:met}
\end{equation}
For  axially symmetric  fields 
 $\psi$  is function  of the coordinates $R$ and $z$
only. The  Einstein-Maxwell equations, in  geometrized units such that 
$ G = c  = 1$,  are given by 
\begin{subequations}\begin{eqnarray}
   R_{ab} \  &  = &  \ 8 \pi (T^{ {\text mat} }_{ab} +  T^{ {\text em} }_{ab}  ), \label{eq:einmax1}  \\
\nabla_b F^{ab} &= &  4 \pi   J^{a},  \label{eq:einmax2}
\end{eqnarray}\end{subequations} 
where
\begin{subequations}\begin{eqnarray}
 T_{ab} \  & = & \ \frac{1}{4 \pi} \left [ F_{ac}F_b^{ \ c} - \frac 14
g_{ab}F_{cd}F^{cd} \right ]   , \label{eq:tab}   \\
F_{ab} &=&  A_{b,a} -  A_{a,b} .
\end{eqnarray}\end{subequations} 
The other symbols have the usual meaning, i.e.,  $( \ )_{,a}=\partial /\partial
x^a$,  $\nabla_b$   covariant derivate, etc. 

For a charged dust distribution  
\begin{subequations}\begin{eqnarray}
 T^{ {\text mat} }_{ab}  &  = &   \rho V_a V_b,   \\
A_a  & =  & \delta ^0 _a \phi ,
\end{eqnarray}\end{subequations} 
where $\rho$ is the  energy density, $V^a$ the four-velocity of the fluid,
and $\phi$ electrostatic potential, 
 the Einstein-Maxwell equations have solution  in the case of  spacetime (\ref{eq:met}) \cite{majum,papa,bonnor98}
 \begin{subequations}\begin{eqnarray}
\phi &=& k (1-\psi)^{-1} , \label{eq: phi} \\
 \rho & = &  \frac { \nabla ^2 \psi } {  4 \pi  (1-\psi)^3} ,  \label{eq: rho}   \\
\sigma &=& k \rho, \label{eq: sigma}
\end{eqnarray}\end{subequations} 
 where $\sigma$ is the electric charge density and $k=\pm 1$.
 $V^a$ is the timelike vector 
of the orthonormal  tetrad  (comoving observer) ${{\rm e}_{ (a)}}^b = \{
V^b , W^b , X^b,Y^b \}$, where
\begin{subequations}\begin{eqnarray}
V^a &=&   (1-\psi)   \delta ^a_t  , \quad   \quad 	
W^a =  (1-\psi)^{-1}  \delta ^a_\varphi / R , \label{eq:tetrad1}	\\
X^a &=&  (1-\psi)^{-1}  \delta ^a_R , \quad
Y^a =  (1-\psi)^{-1}  \delta ^a_z.  \label{eq:tetrad2}
\end{eqnarray}\end{subequations}

 In the Newtonian limit  $\psi \ll 1$,   the expression 
relativistic  (\ref{eq: rho}) reduces to the  Poisson's
equation and  in absence of matter to the Laplace's equation.

Exact solutions  of  Einstein-Maxwell's equations representing the field of a 
disk   of infinitesimal thickness  immersed or not in a matter distribution  can be obtained   applying  a  Kuzmin-Toomre  transformation  \cite{Kuzmin,Toomre}
 $z \rightarrow a + |z|$,
where  $a$ is  a    positive constant, and models of thick disks using   a  Miyamoto-Nagai transformation  which consists in changing   in the  Kuzmin-Toomre thin disks  $ |z| \rightarrow \zeta = \sqrt{z^2 + b^2} $, where $b$ is a positive  parameter.
When this  procedure is applied to a electrovacuum  Einstein-Maxwell solution   
 (\ref{eq: phi}) - (\ref{eq: sigma}), we obtain a relativistic charged dust thick disk with
energy density given by 
\begin{equation}
\rho  =  \frac{  b^2 \left ( \zeta^ {-1}\psi_{, \zeta }  -  \psi _{,\zeta \zeta } \right ) }  {4 \pi  \zeta ^2  (1-\psi)^3}.
\end{equation}

A useful  parameter related  to the motion of test  particles around the structures on the
equatorial plane  is circular speed $v_c$  (rotation curves).
For circular, equatorial  orbits the 4-velocity  ${\bf u}$  of the particles with respect to the coordinates
frame has  components ${\bf u} = u^t(1,\omega , 0,0)$, where  $\omega= u^\varphi/u^t=\frac{d
\varphi}{d t}$ is  the  angular speed of the test particles. 
With respect to tetrad (\ref{eq:tetrad1})  -  (\ref{eq:tetrad2}),  the 4-velocity  has component
\begin{equation}
u^{(a)} = e^{(a)}_{\ \ \ b} u^b,
\end{equation}
and  the  3-velocity 
\begin{equation}
v^{(i)} = \frac { u^{(i)} } { u^{(t)} }  =   \frac { e^{(i)}_{\ \  \ a} u^a }{  e^{(t)}_
{ \ \ \ a}  u^a  }. 
\end{equation}

For circular, equatorial  orbits   the
only nonvanishing velocity  component is $v^{ (\varphi)}$, and is given by
\begin{equation}
 [v^{ (\varphi)}]^2 = v_c^2= - \frac{ g_{\varphi \varphi} }{ g_{tt} } \omega ^2 ,  \label{eq:vc2}
\end{equation}
and  represents   the circular speed   (rotation profile)   of the particle as seen by an observer at infinity.

The angular speed  $\omega$ can be calculated  considering  the electrogeodesic motion of the particles.  For the spacetime   (\ref{eq:met}), the radial motion's equation  is given by 
\begin{equation}
\frac 12   g_{ab, R}u^a u^b = - \tilde e  F_{R a} u^a_,
\label{eq:geo}
\end{equation}
where $\tilde e$ is the specific electric  charge of the particles.  For the  electrostatic case we have 
\begin{equation}
 \frac 12 u^0 (g_{\varphi \varphi, R } \omega ^2 + g_{t t, R} )= - \tilde e  \phi_{,R }, \label{motion}
\end{equation}
where $u^0$ obtains normalizing  $u^a$, that is requiring  $g_{ab}u^au^b=-1$, so that
\begin{equation}
 (u^0)^2 = - \frac {1}{g_{\varphi \varphi} \omega^2 + g_{tt}} \label{eq:u0}.
\end{equation}

Thus, the angular speed  $\omega $  is given by
\begin{equation}
\omega ^2 = \frac { - T_1 \pm \sqrt { T_1^2-4T_2 T_3 }}{ 2  T_2 }, \label{eq:omega}
\end{equation}
where
\begin{subequations}\begin{eqnarray}
T_1 & = & 2 g_{t t, R}  g_{\varphi \varphi, R} +  4 \tilde e ^2 g_{ \varphi \varphi }  \phi_{, R}^2,    \\
T_2 & = &  g_{\varphi \varphi, R}^2   ,  \\
T_3 & =& g_{t t, R}^2  +  4 \tilde e ^2 g_{ tt }  \phi_{, R}^2.
\end{eqnarray}\end{subequations}
The  positive sign corresponds to the direct orbits or co-rotating
and the negative sign  to the retrograde orbits or counter-rotating.

To  analyze the stability of the particles against radial
perturbations we can use an extension of the Rayleigh criteria of stability of a fluid in rest in a
gravitational field 
\begin{equation}
\frac{d(h^2)}{dR} \ > \ 0 ,
\end{equation}
where $h$  is the specific
angular momentum,   defined as $h =
g_{\varphi a} u^a$. For  circular, planar orbits  we obtain 
\begin{equation}
h^2 = \frac{R^2 \left( 1- \psi \right)^{2}  v_c ^2}{1- v_c ^2} .
\label{eq:h}
\end{equation}
All above quantities are evaluated on the equatorial plane $z = 0$.

\section{ Charged dust thick disks from a extreme  Reissner-Nordstr\"om  black hole } 

For a extreme  Reissner-Nordstr\"om black hole 
\begin{equation}
\psi _{BH}= - \frac{M} {\sqrt {R^2 + z^2}} 
\end{equation}
and  the application of  the Miyamoto-Nagai transformation yield the potential 
\begin{equation}
\psi_D = -\dfrac{M}{\sqrt{R^2 + \left(a +\sqrt{ z^2 + b^2}\right)^2}} ,
\end{equation}
which corresponds  in Newtonian gravity to  the  first   Miyamoto-Nagai  model (with $G=1$). The relativistic energy density is given by
\begin{equation}
\tilde \rho  =    \frac{ \tilde b^2
\left[  \tilde a  \tilde R^2 + ( \tilde a + 3 \sqrt{ \tilde z^2 + \tilde b^2 }  ) ( \tilde a +  \sqrt{\tilde z^2 + \tilde b^2 } )^2 \right]   }
{ 4\pi \left[\tilde  R^2+\left( \tilde a+\sqrt{\tilde z^2 + \tilde b^2 }\right) ^2 \right]        \left [  1 + \sqrt{\tilde R^2 + \left(\tilde a +\sqrt{ \tilde z^2 + \tilde b^2}\right)^2}     \right]^3 ( \tilde z^2 + \tilde b^2 )^{3/2} } ,
\end{equation}
where $\tilde \rho = M^2 \rho $,  $\tilde R = R / M $,
 $\tilde z = z / M $,     $\tilde a = a / M $  and  $\tilde b = b / M $.
Note that the energy density is always  
a positive quantity  in agreement with  the  weak energy condition.

\begin{figure}
$$
\begin{array}{cc}
\includegraphics[width=0.33\textwidth]{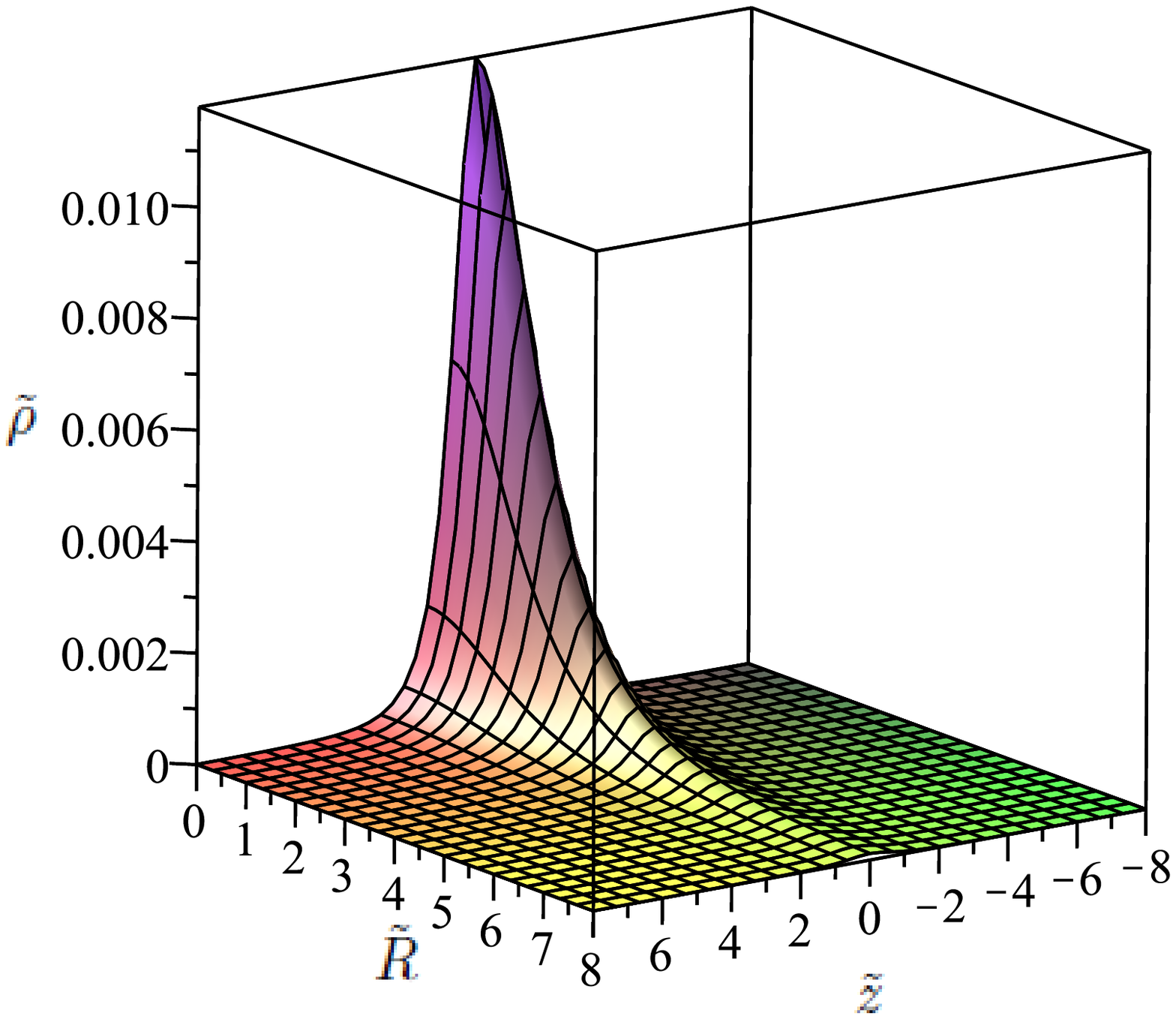} &   
\includegraphics[width=0.25\textwidth]{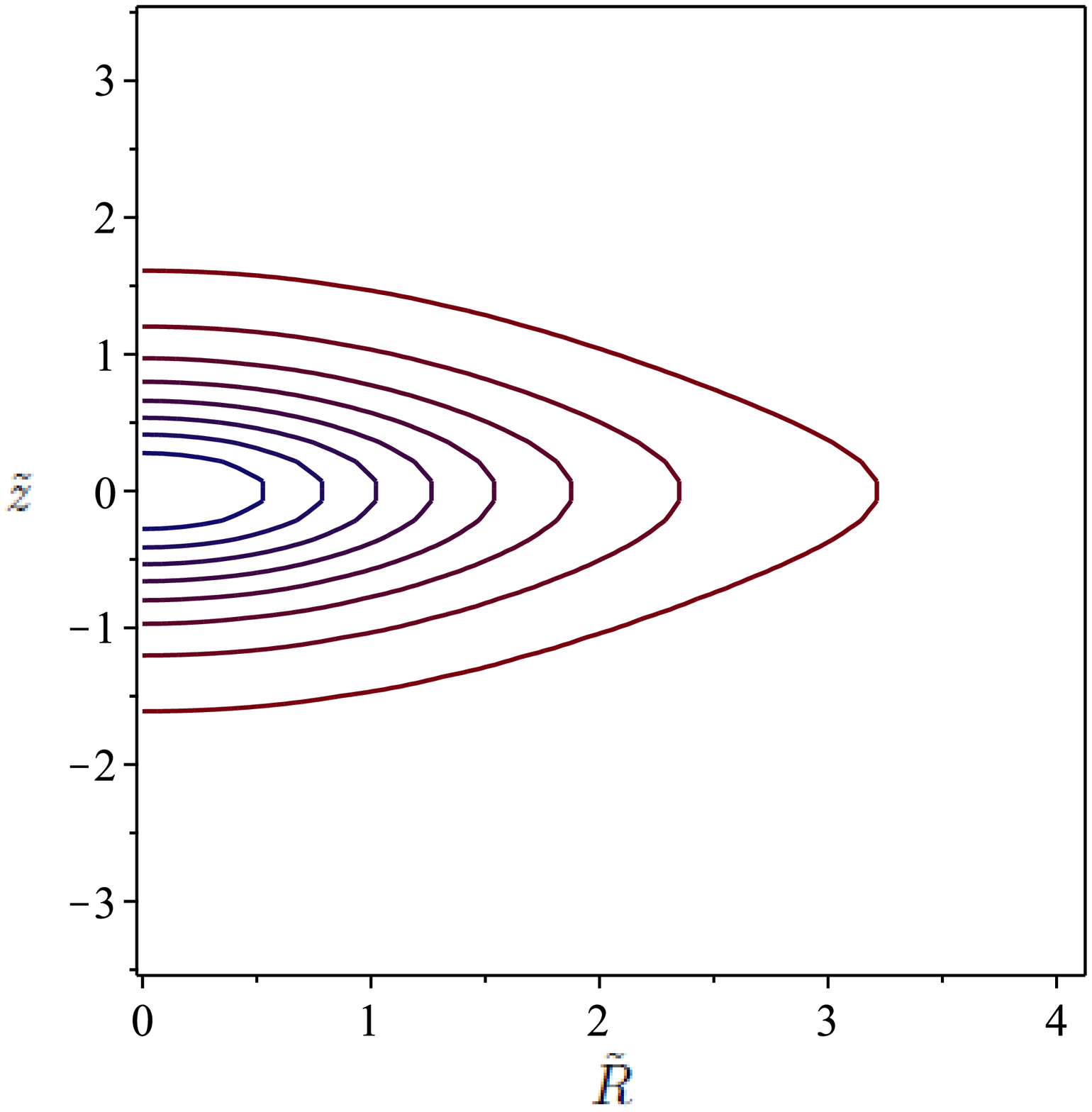}  \\  
 &  \\
(a)    &  (b)  \\
\includegraphics[width=0.33\textwidth]{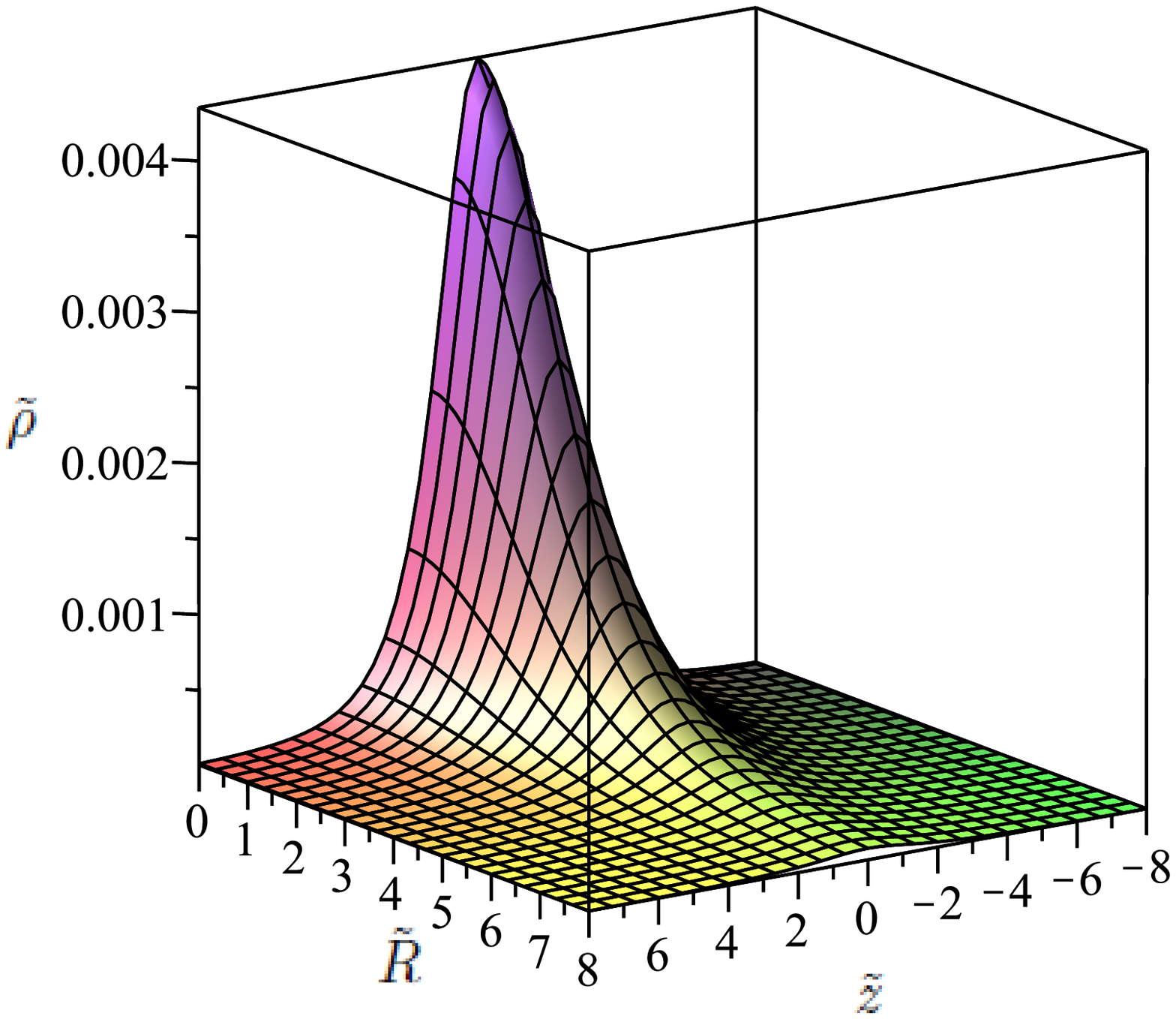} &   
\includegraphics[width=0.25\textwidth]{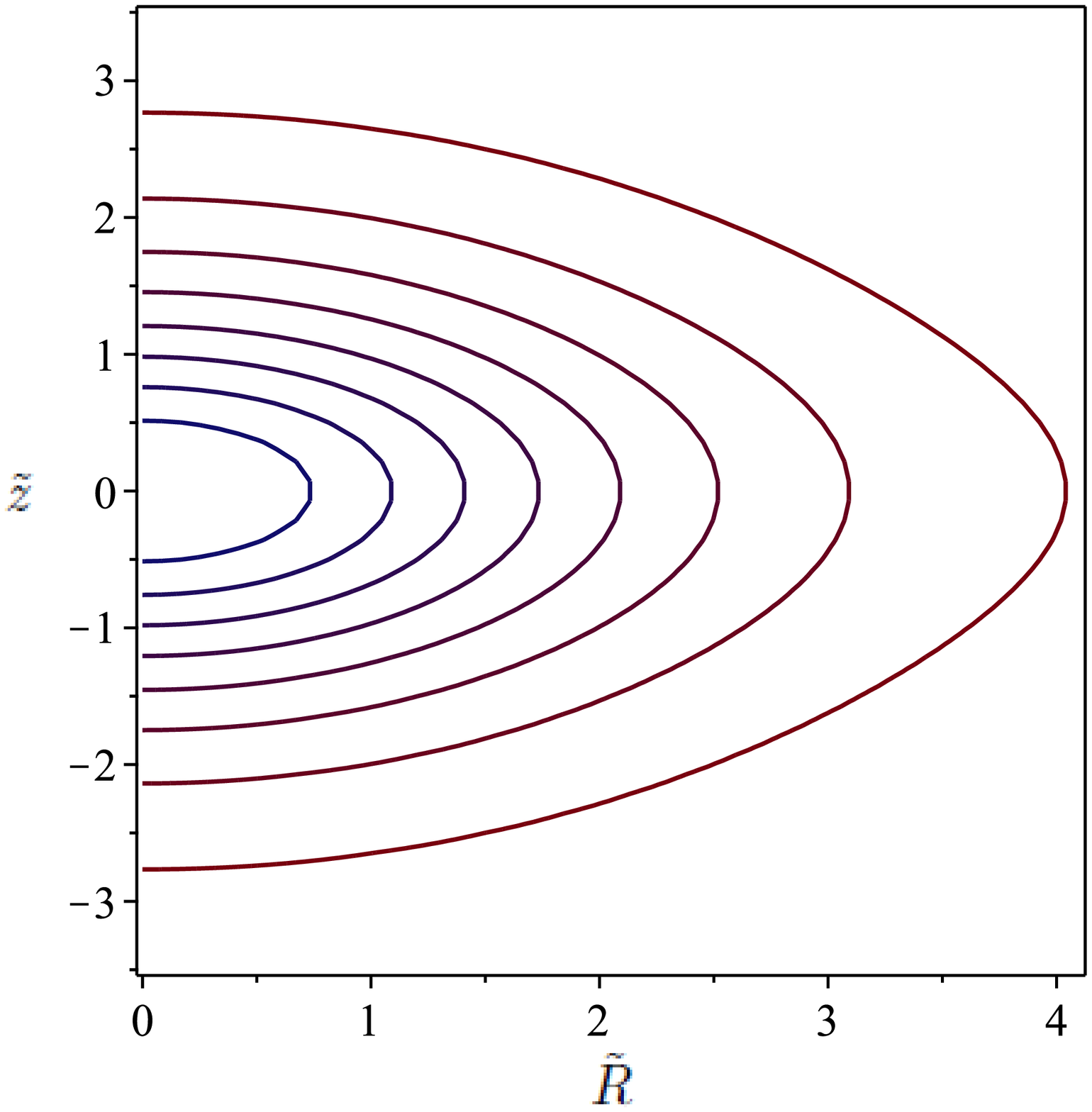}  \\  
 &  \\
(c)    &  (d)
\end{array}
$$	
\caption{The  relativistic energy density $\tilde \rho$ and contour plots for  extreme  Reissner-Nordstr\"om-like charged dust thick disks 
with parameters
 $\tilde a = 1$, $\tilde b =  1$ (top figures) and  $\tilde b =  2$,   as functions of  $\tilde R$ and $\tilde z$. } 
\label{fig:fig1}
\end{figure}

In Figure  \ref{fig:fig1}  we graph the relativistic  energy density  $\tilde \rho$   and the contour curves   for extreme  Reissner-Nordstr\"om-like  charged dust thick disks 
with parameters
 $\tilde a = 1$  ($a=M$),  $\tilde b =  1$ and  $ 2$,   as functions of 
$\tilde R$ and $\tilde z$. 
The energy density  presents a   maximum value on $\tilde R=0$  
and  then decreases rapidly with $\tilde R$    which  permits
to define a cut off radius $\tilde R_c$ and, in principle, to model 
these matter distributions  as  compact objects. Furthermore, as in the  Newtonian case,  as the ratio $ b /  a$
decreases the distribution of energy  becomes flatter so that $b / a $ is also  a 
measure of flatness of the models.

\begin{figure}
$$
\begin{array}{cc}
v_+^2   &  v_-^2  \\  
\includegraphics[width=0.3\textwidth]{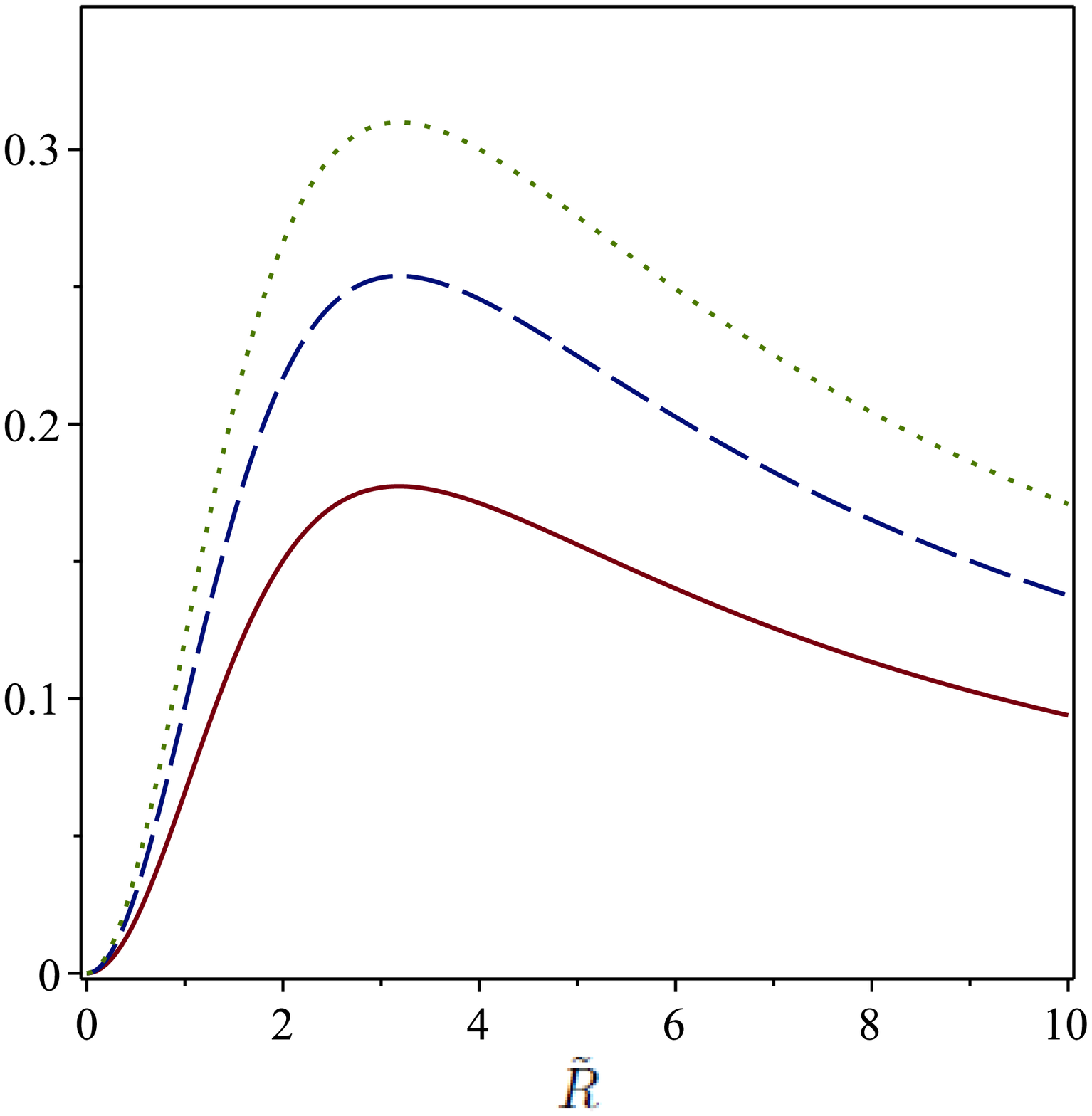} &   
\includegraphics[width=0.3\textwidth]{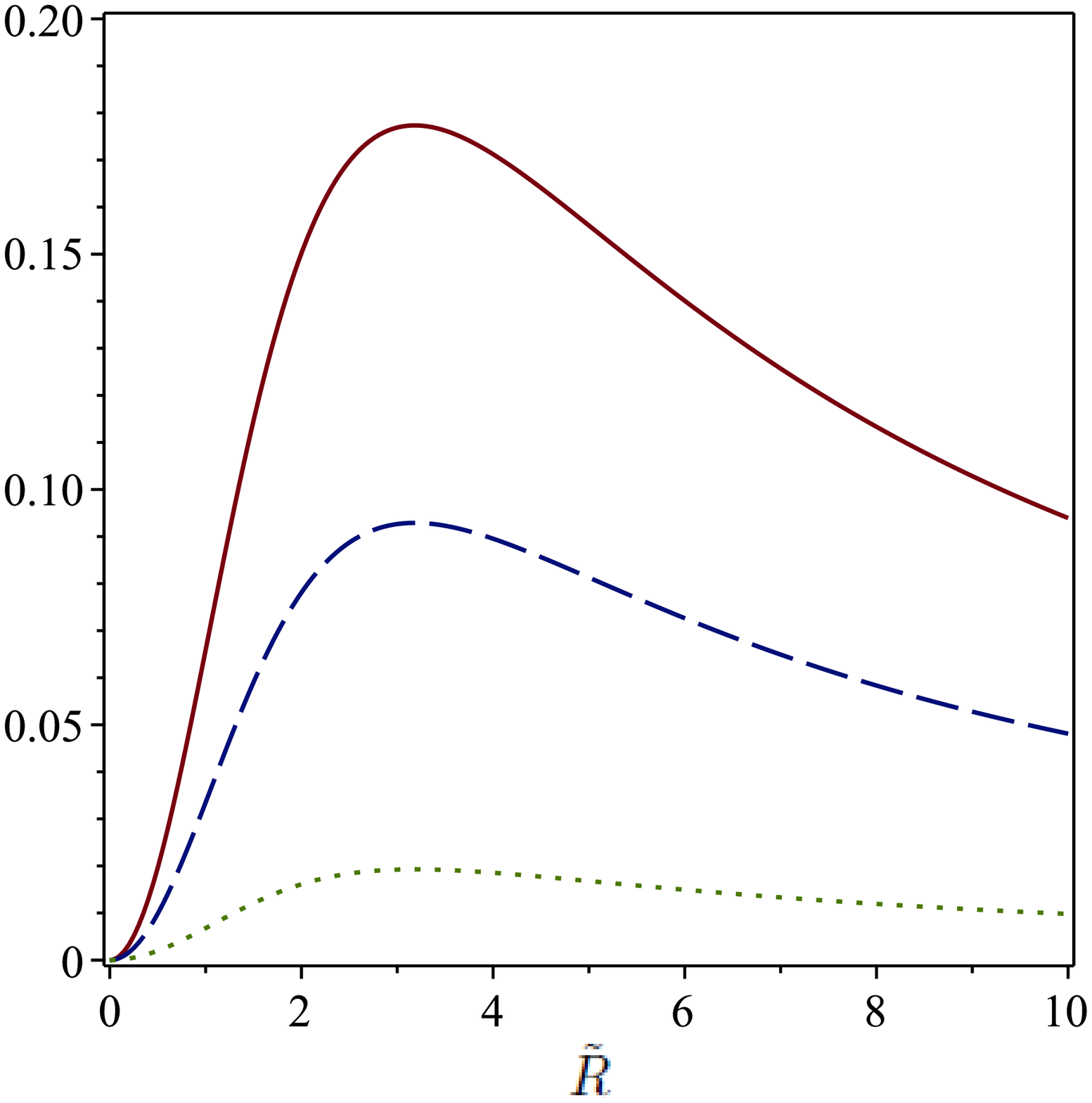}  \\  
 &  \\
(a)    &  (b)  \\
\includegraphics[width=0.3\textwidth]{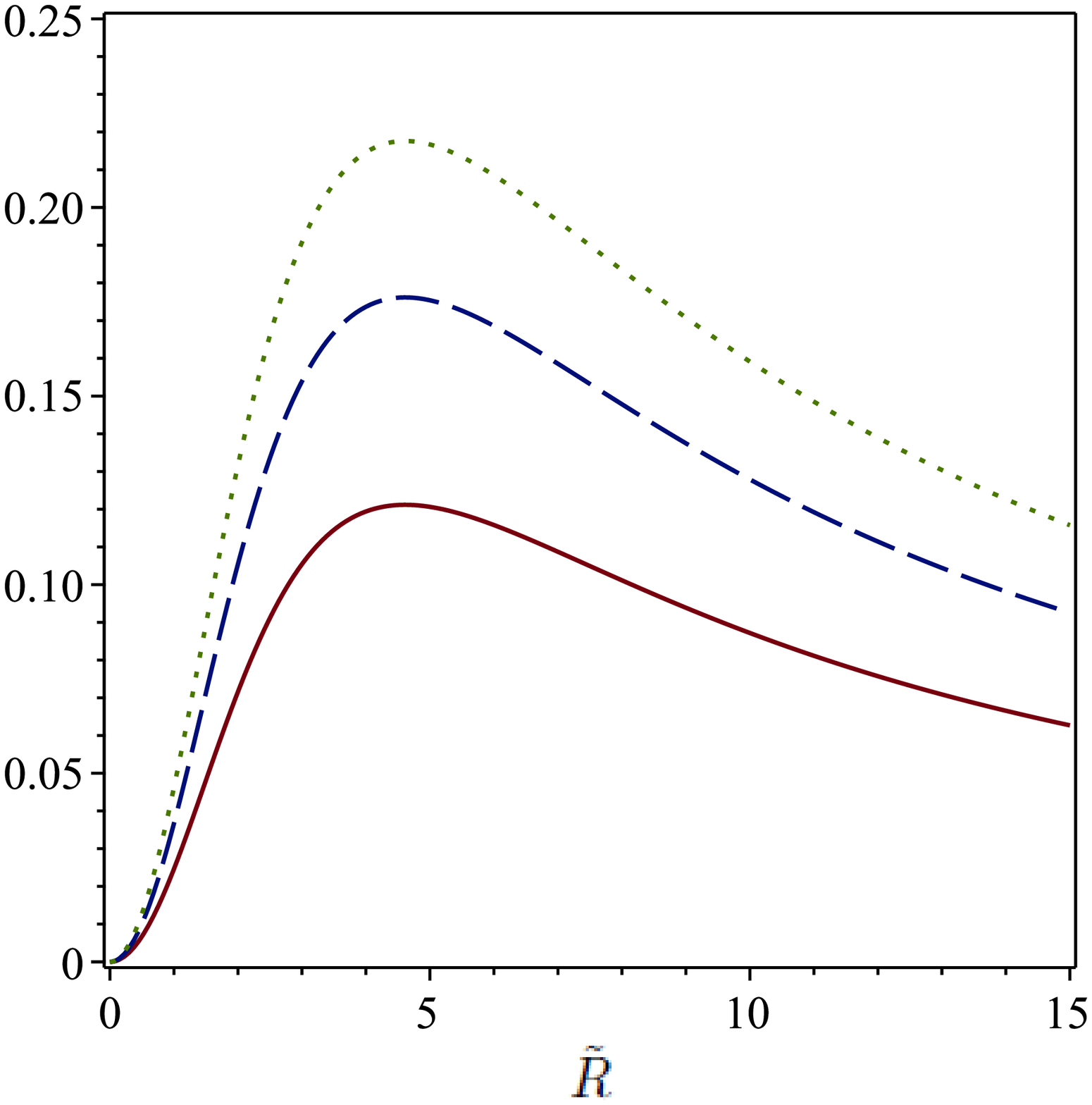} &   
\includegraphics[width=0.3\textwidth]{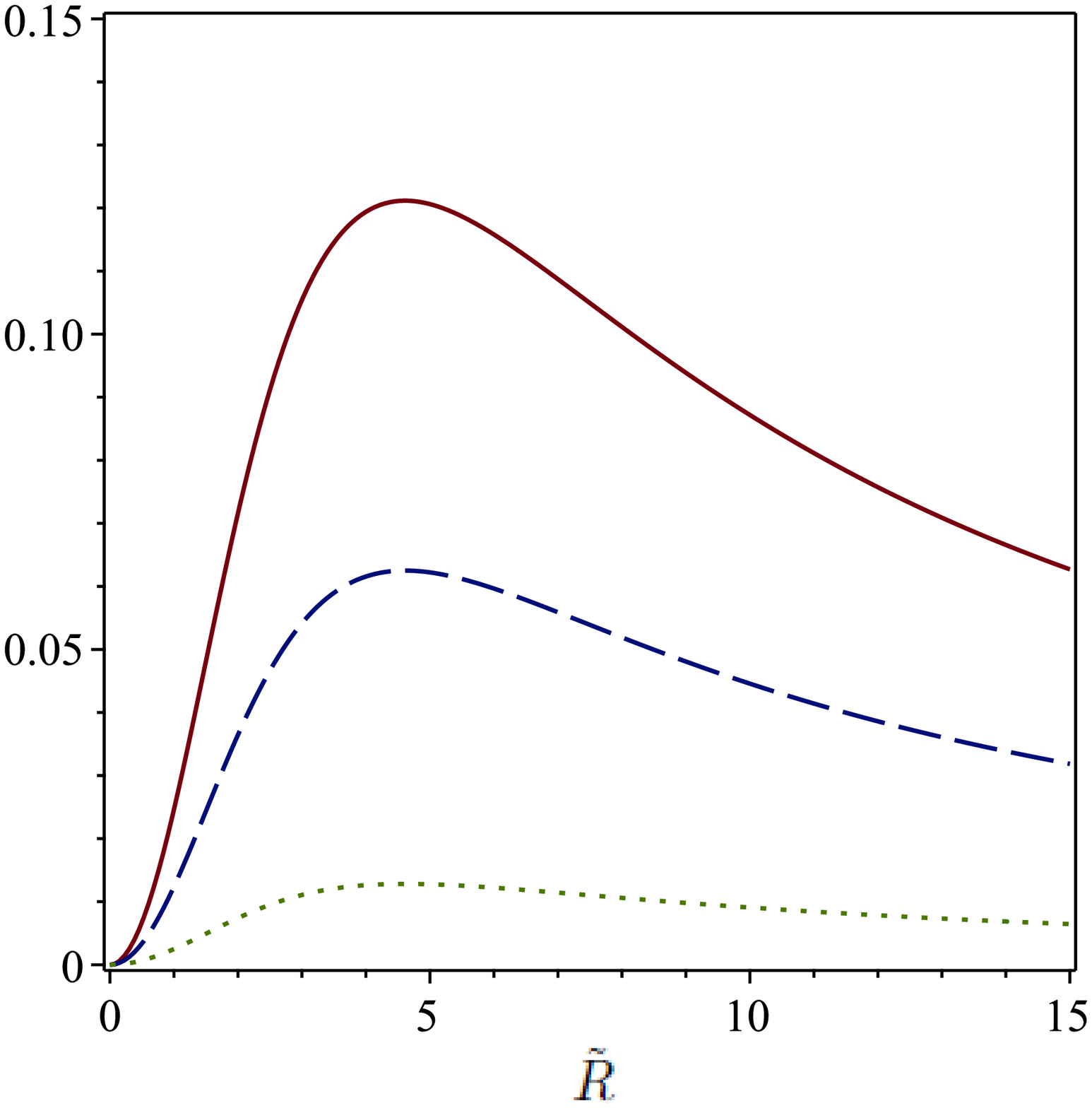}  \\  
 &  \\
(c)    &  (d)
\end{array}
$$	
\caption{ The
circular speed  $v^2_+$ and  $v^2_-$ for  charged test particles
around  extreme  Reissner-Nordstr\"om-like charged  dust thick disks with parameters  $\tilde e = 0$  (solid curves),  $0.5$,  $0.9$ (dotted curves),  $\tilde a = 1$,   $\tilde b =  1$ (top figures) and   $\tilde b =  2$,   as functions of 
$\tilde R$   } 
\label{fig:fig2}
\end{figure}

\begin{figure}
$$
\begin{array}{cc}
h_+^2   &  h_-^2  \\  
\includegraphics[width=0.3\textwidth]{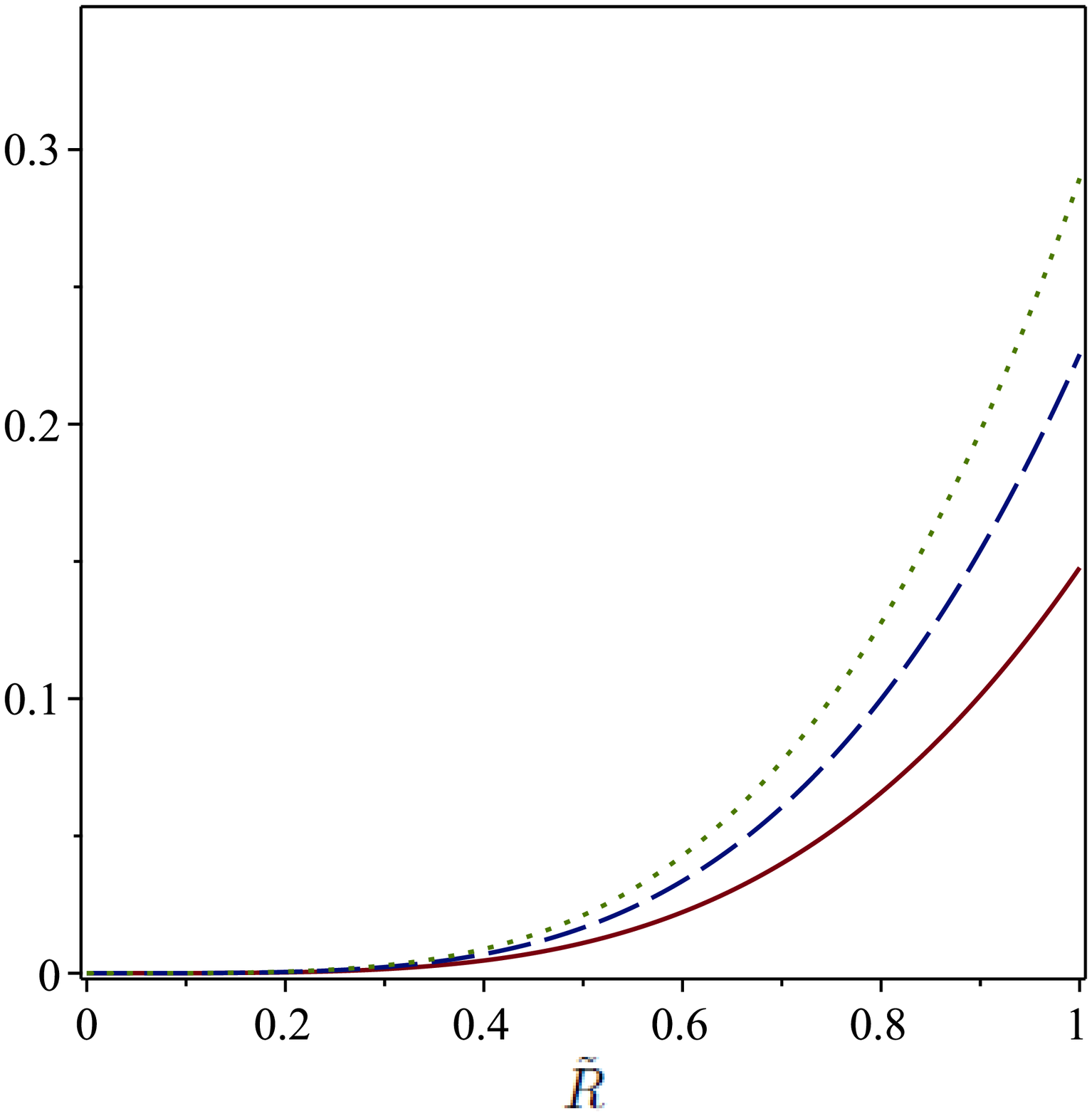} &   
\includegraphics[width=0.3\textwidth]{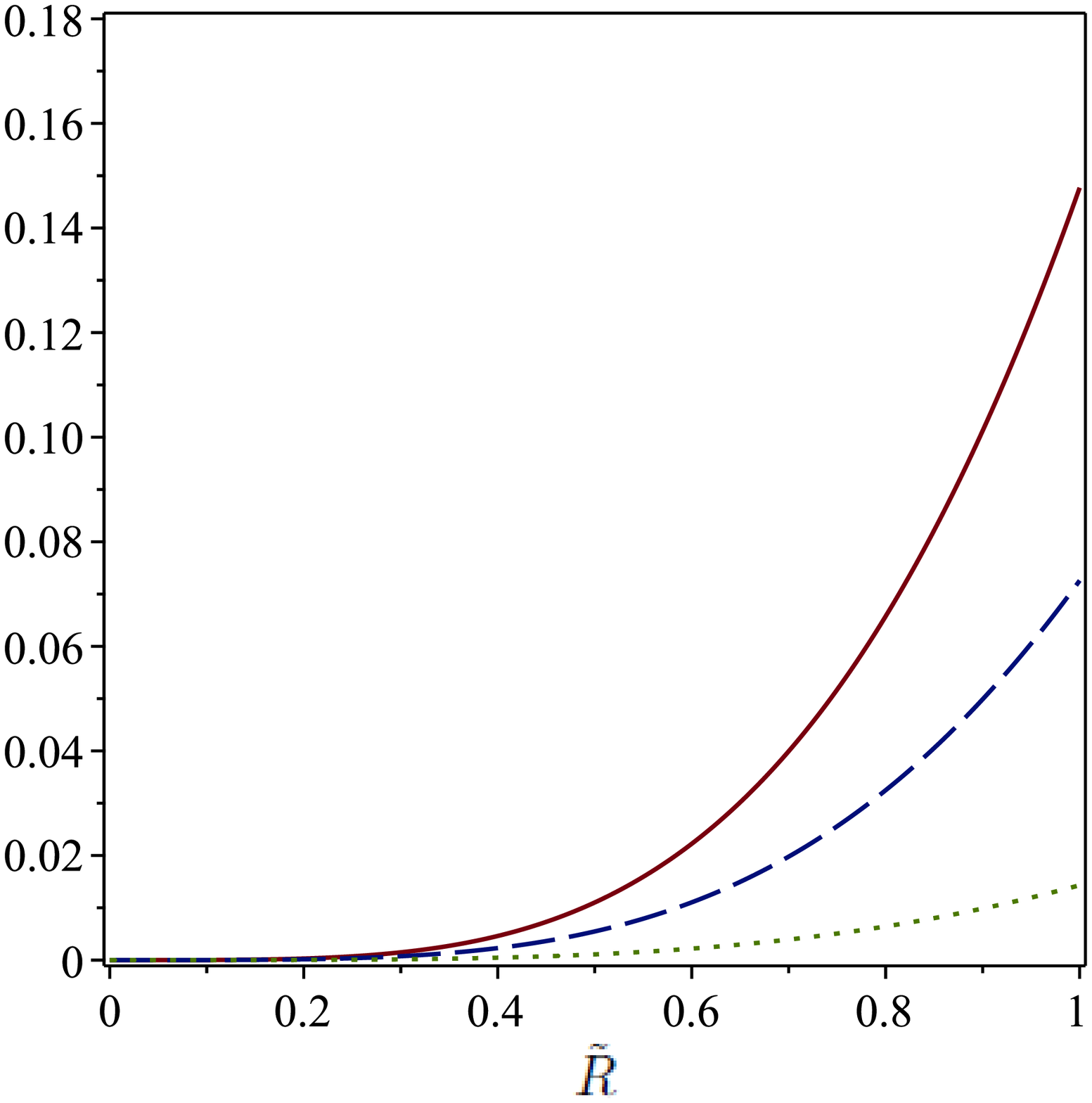}  \\  
 &  \\
(a)    &  (b)  \\
\includegraphics[width=0.3\textwidth]{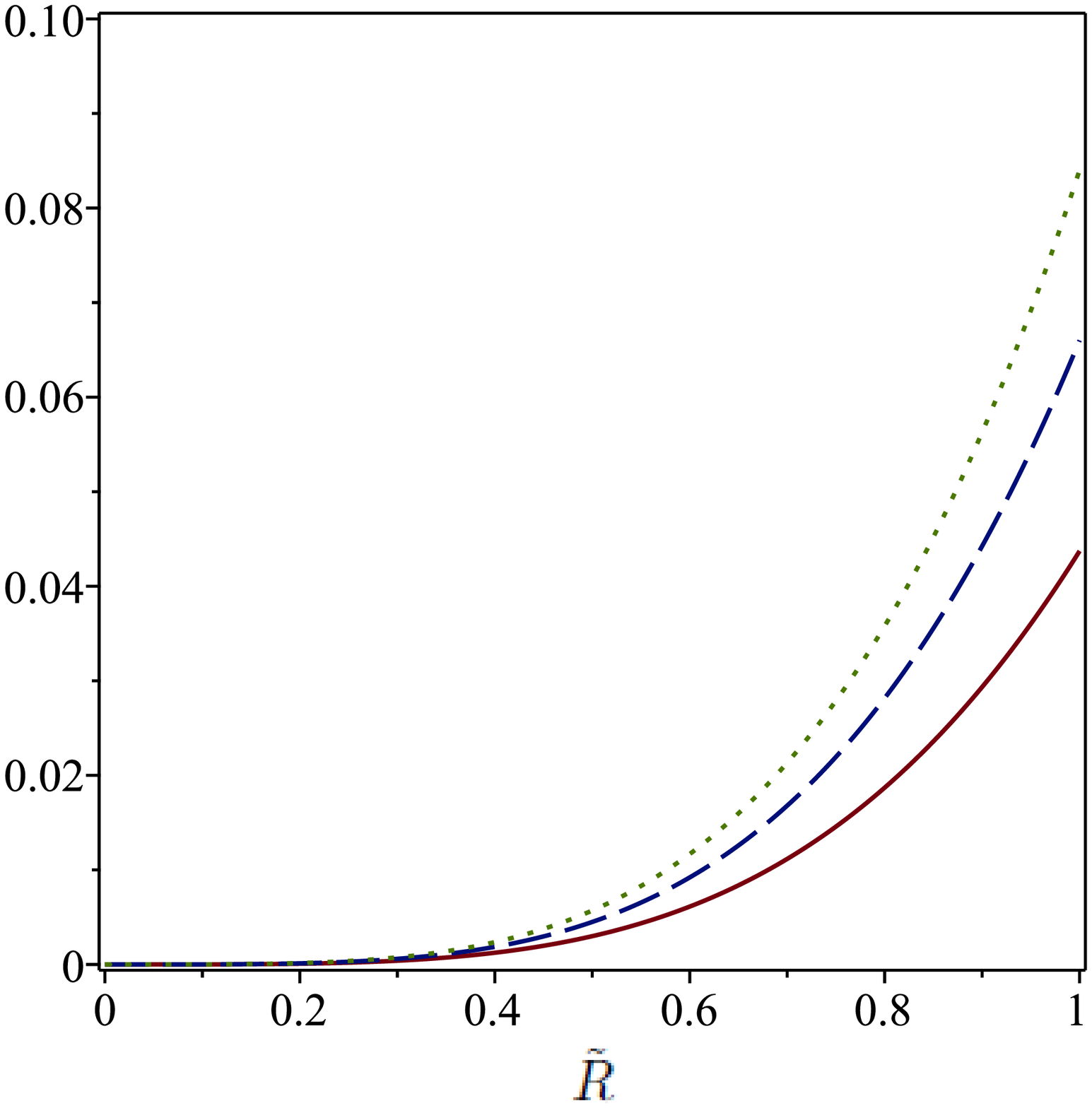} &   
\includegraphics[width=0.3\textwidth]{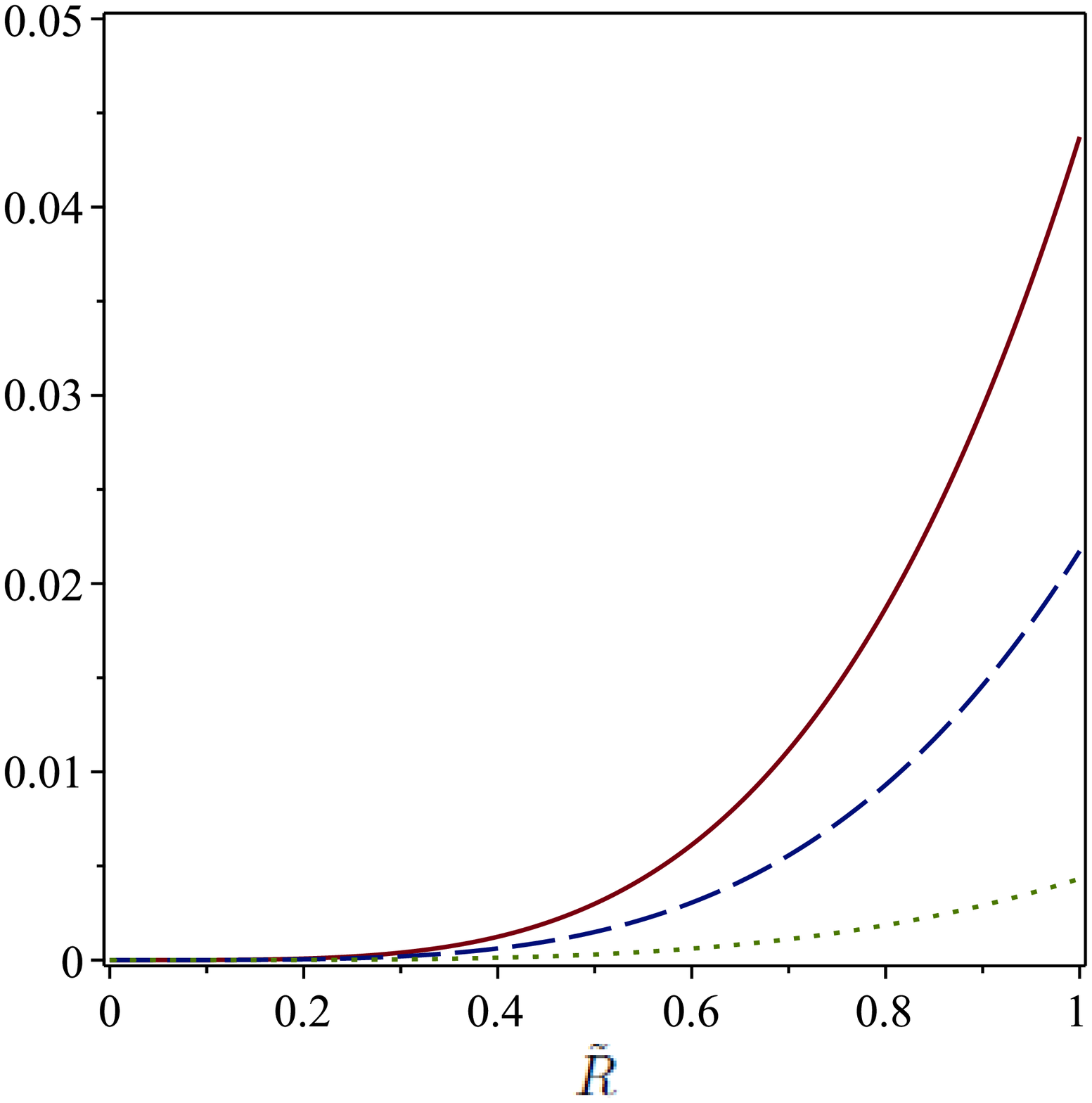}  \\  
 &  \\
(c)    &  (d)
\end{array}
$$	
\caption{The
specific angular momentum   $h^2_+$ and  $h^2_-$ for  charged test particles
around  extreme  Reissner-Nordstr\"om-like charged  dust thick disks with parameters  $\tilde e = 0$  (solid curves),  $0.5$,  $0.9$ (dotted curves),  $\tilde a = 1$,   $\tilde b =  1$ (top figures) and   $\tilde b =  2$,   as functions of 
$\tilde R$ } 
\label{fig:fig3}
\end{figure}

In figure  \ref{fig:fig2} we illustrate the behavior of the
circular speed for direct orbits $v^2_+$ and  retrograde orbits  $v^2_-$ of  charged test particles
with  $\tilde e = 0$ (neutral particles),   $0.5$, $0.9$ and the same values of the parameters $\tilde a$ and  $\tilde b$.
We find that for direct orbits  the specific electric charge  increases the speed of the particles whereas
for retrograde orbits the contrary occurs. Also,  for direct orbit the particles  become more relativistic  when the matter distribution is flatted, 
whereas
for retrograde orbits   the opposite  occurs.
We observer  that  circular speed  of particles is always a quantity
less than the speed  of light in agreement  with the dominant energy condition. 
In figure \ref{fig:fig3}   we also  show,  as function of $\tilde R$,  the specific angular momentums $\tilde h^2_+$ and  $\tilde h^2_-$  for  the same values of parameters. We find that for these values  the orbits of the particles are  stable against radial perturbation.  

\section{Morgan-Morgan-like  charged dust  thick disks } 

Thin disks of finite extension can be obtained in Newtonian gravity    solving
the Laplace equation in oblate  spheroidal coordinates
($u$,$v$), which are defined in terms of the cylindrical
coordinates  ($R$, $z$)  by
\begin{subequations}\begin{eqnarray}
 R^2 & =& d ^2 (1 + u^2 )(1 - v^2 ) ,  \\
 z & = & duv,
\end{eqnarray}\end{subequations}
and explicitly
\begin{subequations}\begin{eqnarray}
 \sqrt 2 u  & = &  \sqrt { [ ( \tilde R^2 + \tilde z^2 -1)^2 + 4\tilde z ^2
]^{1/2} +  \tilde R^2 + \tilde z^2 -1 } ,  \\
 \sqrt 2 v  & = & \sqrt { [ ( \tilde R^2 + \tilde z^2 -1)^2 + 4\tilde z ^2
]^{1/2} - ( \tilde R^2 + \tilde z^2 -1) },
\end{eqnarray}\end{subequations}
where $u \geq 0$ , $-1< v <1$, $\tilde R = R/d$ and $\tilde z = z/d$,  being $d$ the radius of the disk.  

In such coordinates, the   general solution of Laplace’s equation  can be written as
\begin{equation}
 \Phi = - \sum\limits_{n = 0}^\infty  {c_{2n} q_{2n} (u)P_{2n} (v)} ,
\label{eq:potenM-M}
\end{equation}
where  $c_{2n}$ are constants, $P_{2n}$ are the Legendre polynomials of order
$2n$ and
\begin{equation}
q_{2n} (u) = i^{2n + 1} Q_{2n} (iu),
\end{equation}
being  $Q_{2n} (iu)$  the Legendre functions of the second kind.
For example, for  the first two  terms in  series (\ref{eq:potenM-M}) ($n = 0$ and $n = 1$),  the gravitational potential    is 
\begin{equation}
\Phi   = -   \frac{M G}{d} \left \{ \cot^{ - 1} (u) + \frac{1}{4}
 \left[ (3u^2  + 1) \cot^{ - 1} (u) - 3u \right] \left( 3v^2  - 1 \right) \right
\},  \label{eq:potenM-M1}
\end{equation}
being $M$  the mass  of the disk.

Note that in general  the potential $\Phi$  is function of $z^2 = |z|^2$. Thus, Newtonian thick disks can  be generated
from the solutions  (\ref{eq:potenM-M})  via  a Miyamoto-Nagai transformation taking  $a=0$.   In particular,  for the potential (\ref{eq:potenM-M1})   we obtain the mass distribution
\begin{equation}
\tilde \rho_N = \frac { 3   \tilde b^2 }{ 4  \pi \tilde d^3
    (x^2 + y^2) x^3   }, 
\end{equation}
where   $\tilde \rho =  M^2 \rho $, $\tilde d = d/ M$,  $\tilde b = b/ d$,  and
\begin{subequations}\begin{eqnarray}
 \sqrt 2 x  & = &  \sqrt { [ ( \tilde R^2 + \tilde \zeta^2 -1)^2 + 4\tilde \zeta ^2
]^{1/2} +  \tilde R^2 + \tilde \zeta^2 -1 } ,  \\
 \sqrt 2 y  & = & \sqrt { [ ( \tilde R^2 + \tilde \zeta^2 -1)^2 + 4\tilde \zeta ^2
]^{1/2} - ( \tilde R^2 + \tilde \zeta^2 -1) },
\end{eqnarray}\end{subequations}
where $ \tilde \zeta = \sqrt{ \tilde z^2 +  \tilde  b^2}$.

The relativistic version of the disk like solutions  (\ref{eq:potenM-M}) was first given in \cite{MM1} and are known in the literature as the  Morgan-Morgan disks.  Likewise, relativistic finite thin disks make of charged dust  can  be constructed taking  the Newtonian gravitational potential  $\Phi$ as  the metric function  $\psi$ and   in consequence
   can  also be  ``fattened'' by applying a Miyamoto-Nagai transformation.  Thus, for the seed potential  (\ref{eq:potenM-M1}) we obtain the relativistic energy density
(with G=1)
\begin{equation}
\tilde \rho =  \frac { 3  \tilde b^2 }{ 4  \pi \tilde d^3
    (x^2 + y^2) x^3   (1-\psi)^3 },  \label{eq: rhoMM}
\end{equation}
where  
\begin{equation}
\psi   = -   \frac{1}{\tilde d} \left \{ \cot^{ - 1} (x) + \frac{1}{4}
 \left[ (3x^2  + 1) \cot^{ - 1} (x) - 3x \right] \left( 3y^2  - 1 \right) \right
\}.  \label{eq: psiMM}
\end{equation}

\begin{figure}
$$
\begin{array}{cc}
\includegraphics[width=0.33\textwidth]{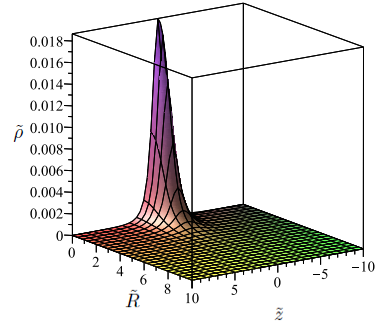} &   
\includegraphics[width=0.25\textwidth]{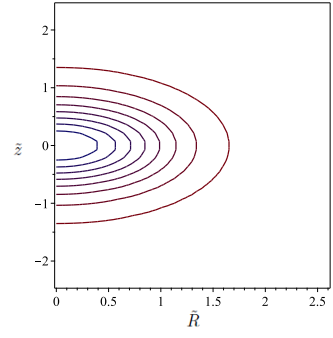}  \\  
 &  \\
(a)    &  (b)  \\
\includegraphics[width=0.33\textwidth]{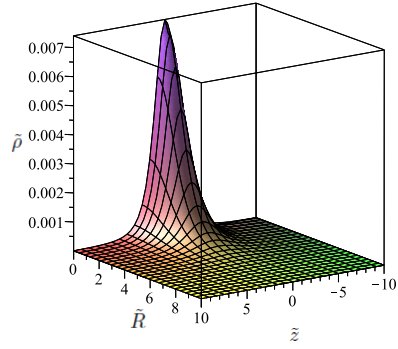} &   
\includegraphics[width=0.25\textwidth]{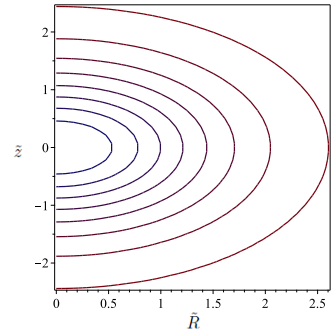}  \\  
 &  \\
(c)    &  (d)
\end{array}
$$	
\caption{The  relativistic energy density $\tilde \rho$ and contour curves for  Morgan-Morgan-like  charged dust thick disks 
with parameters
  $ \tilde d = 1$, $\tilde b =  1$ (top figures) and  $\tilde b =  2$,   as functions of  $\tilde R$ and $\tilde z$. } 
\label{fig:fig4}
\end{figure}

In figure \ref{fig:fig4} we present,  as functions of  $\tilde R$ and $\tilde z$,  the surface and level curves of the 
 energy  density   $\tilde \rho$ for  Morgan-Morgan-like  charged dust thick disks with parameters
$ \tilde d= 1$, $\tilde b =  1$  and  $\tilde b =  2$.  
We see that the energy density is a positive quantity in according with the weak energy condition  and,  unlike  the seed thin disks,
these  disks   have   no  boundary  but as  the distribution of matter   also decreases rapidly with radius $\tilde R$   it    permits to model these structures  as  compact objects.  Since the denominator of  the expression of the energy density is a quantity  greater  than one, the relativistic effects decrease everywhere the energy density.  Here the parameter 
$\tilde b /  \tilde d $ is the
measure of flatness of the models. 

\begin{figure}
$$
\begin{array}{cc}
v_+^2   &  v_-^2  \\  
\includegraphics[width=0.3\textwidth]{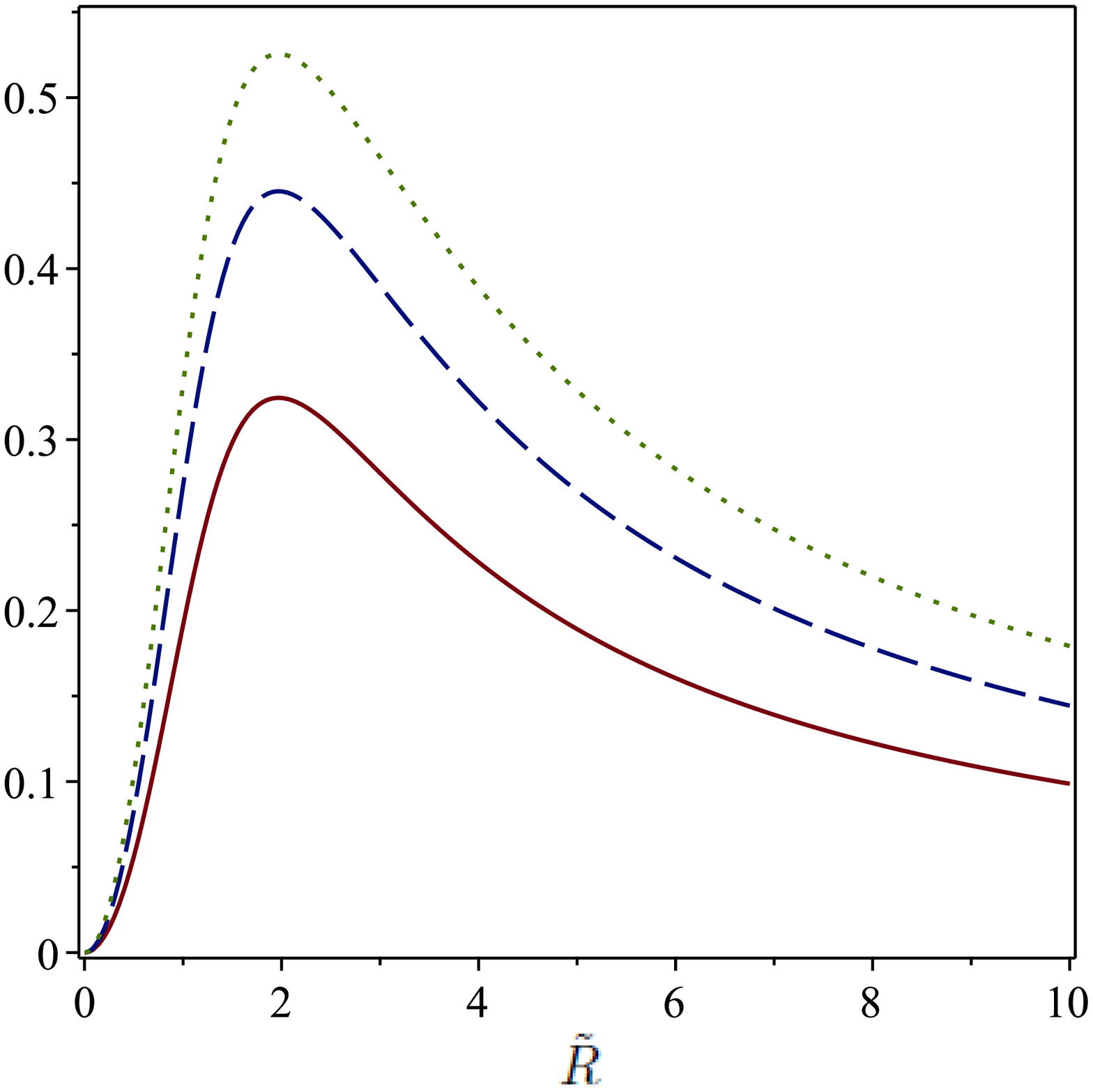} &   
\includegraphics[width=0.3\textwidth]{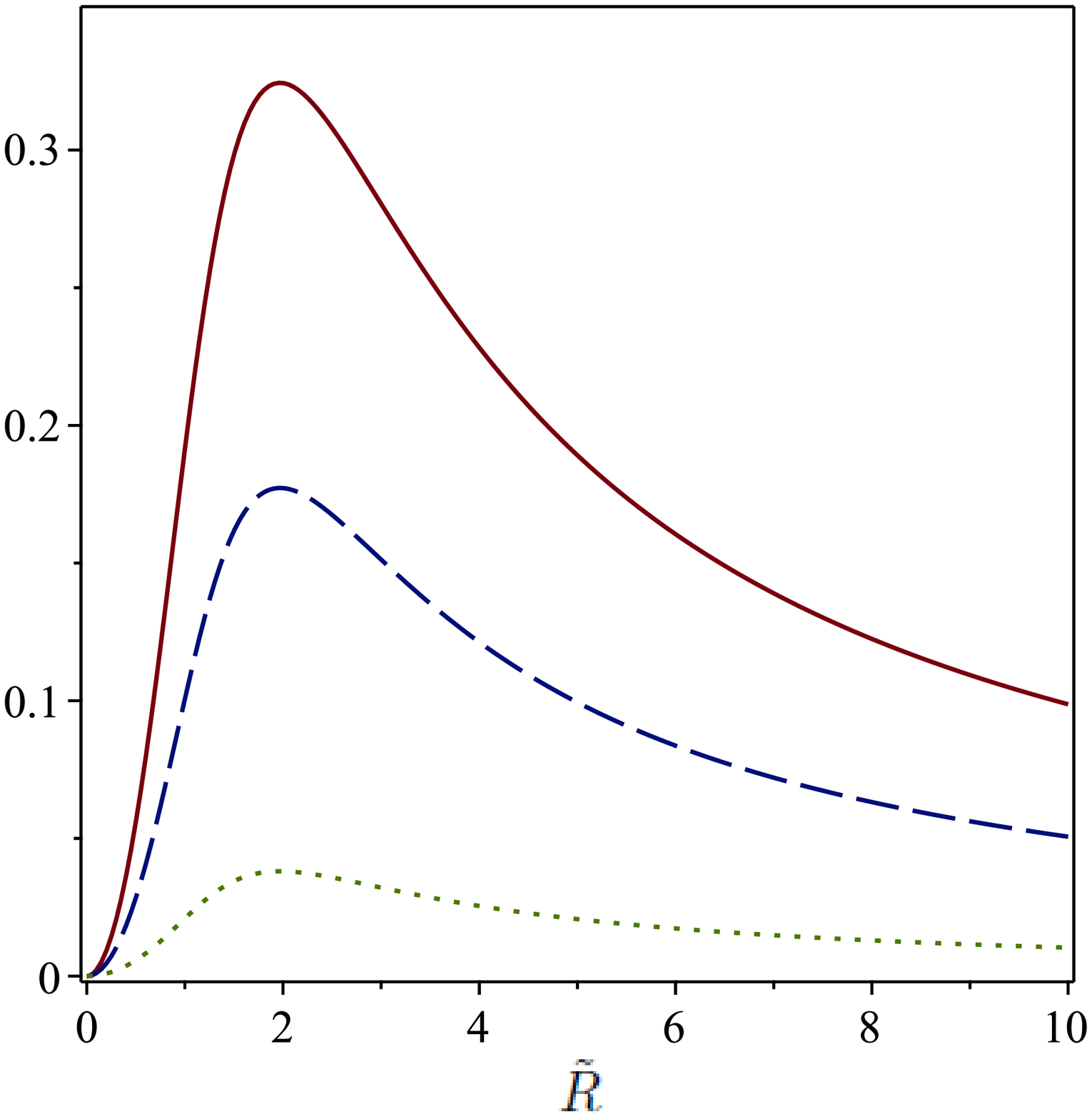}  \\  
 &  \\
(a)    &  (b)  \\
\includegraphics[width=0.3\textwidth]{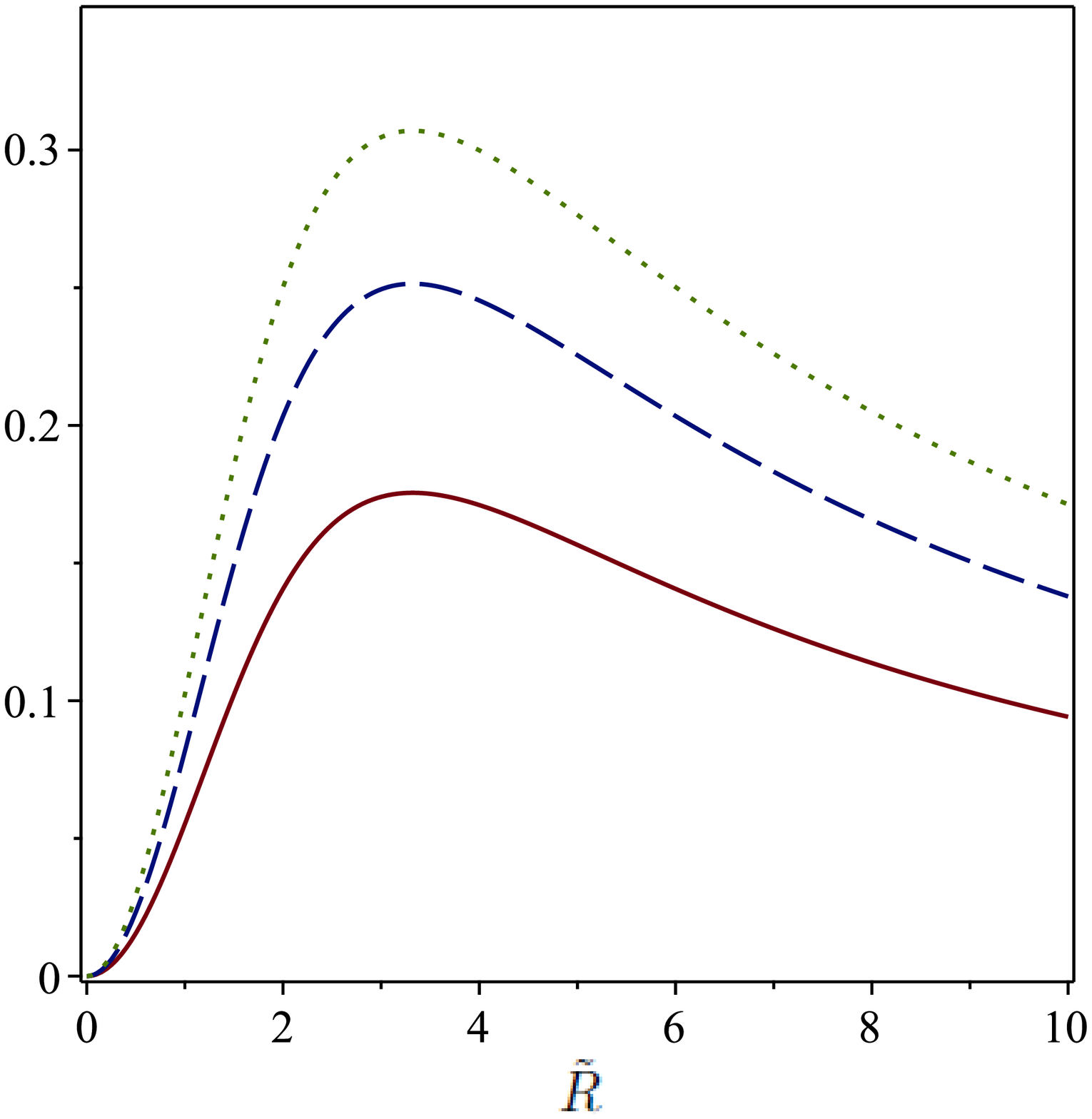} &   
\includegraphics[width=0.3\textwidth]{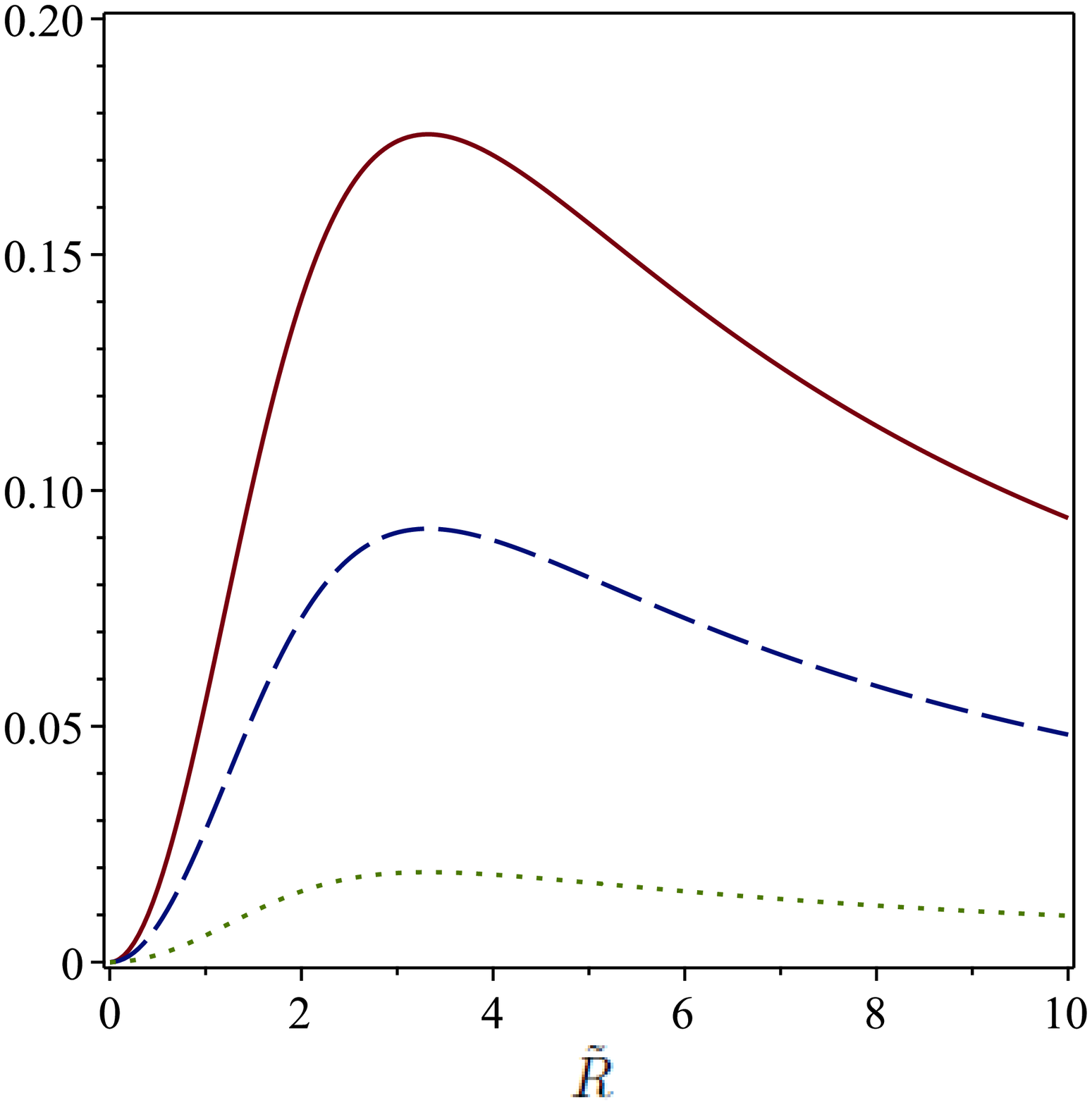}  \\  
 &  \\
(c)    &  (d)
\end{array}
$$	
\caption{ The
circular speed  $v^2_+$ and  $v^2_-$ for  charged test particles
around  Morgan-Morgan-like charged  dust thick disks with parameters  $\tilde e = 0$  (solid curves),  $0.5$,  $0.9$ (dotted curves),   $ \tilde d = 1$,   $\tilde b =  1$ (top figures) and   $\tilde b =  2$,   as functions of 
$\tilde R$   } 
\label{fig:fig5}
\end{figure}

\begin{figure}
$$
\begin{array}{cc}
h_+^2   &  h_-^2  \\  
\includegraphics[width=0.3\textwidth]{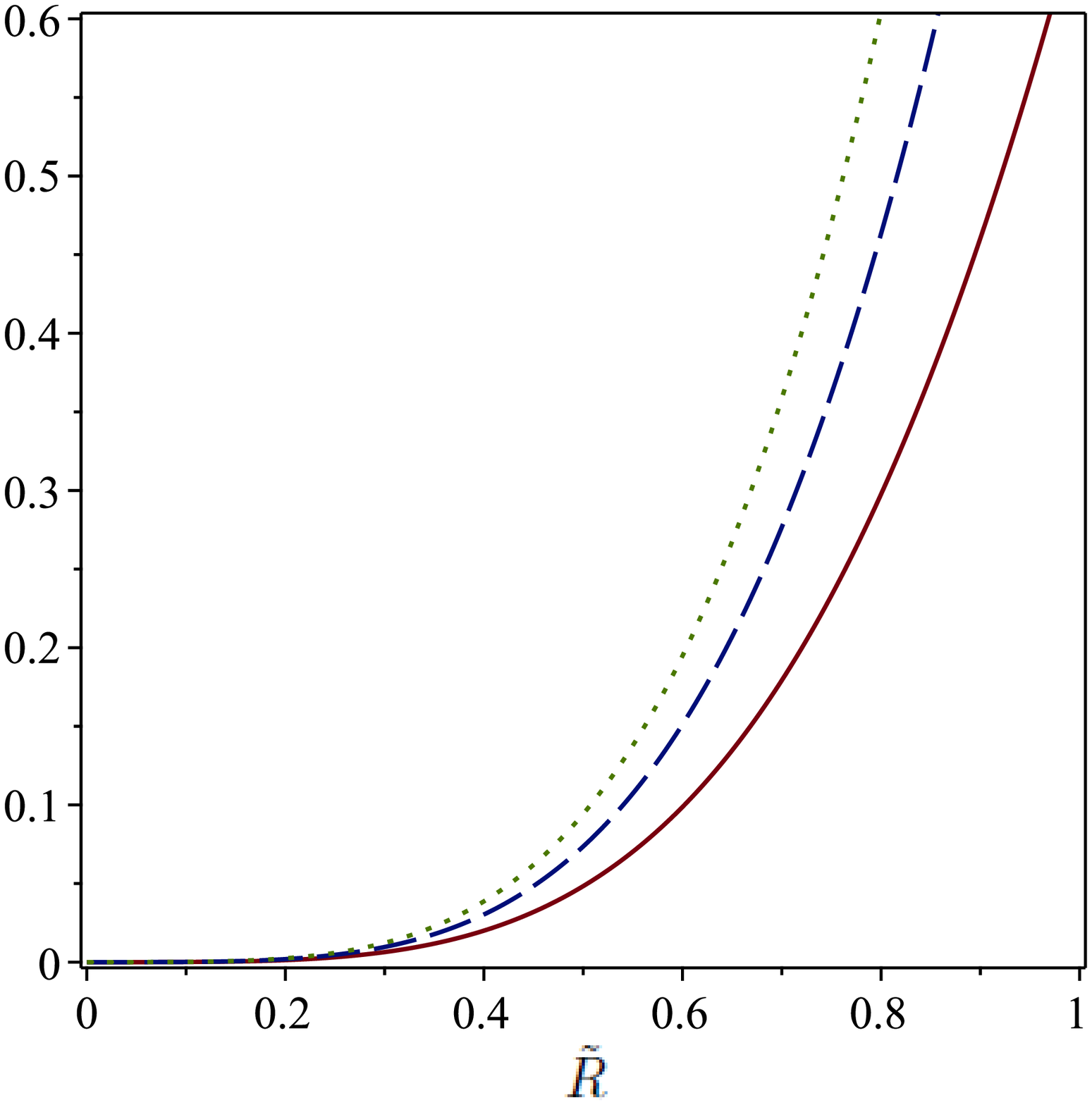} &   
\includegraphics[width=0.3\textwidth]{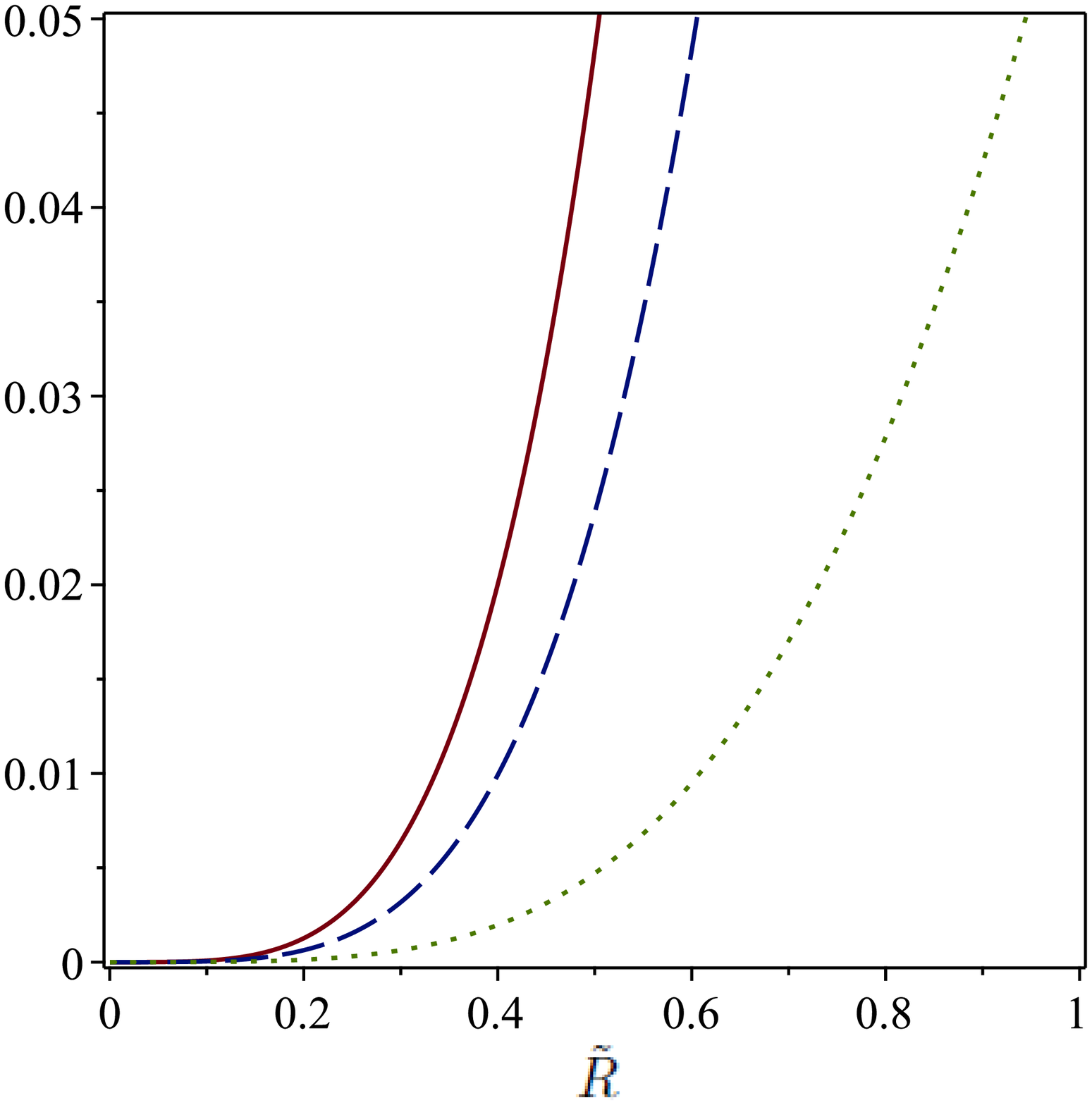}  \\  
 &  \\
(a)    &  (b)  \\
\includegraphics[width=0.3\textwidth]{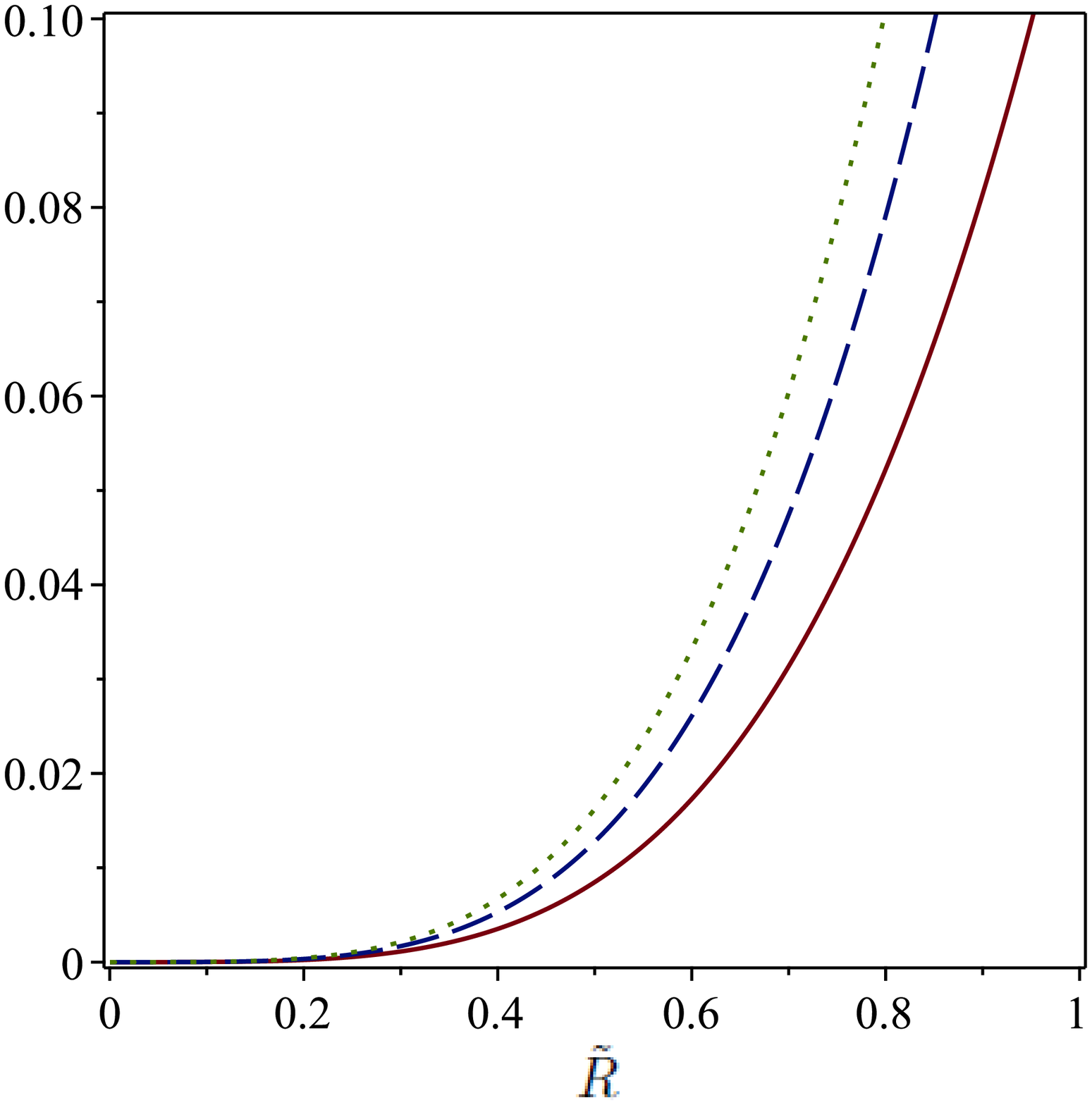} &   
\includegraphics[width=0.3\textwidth]{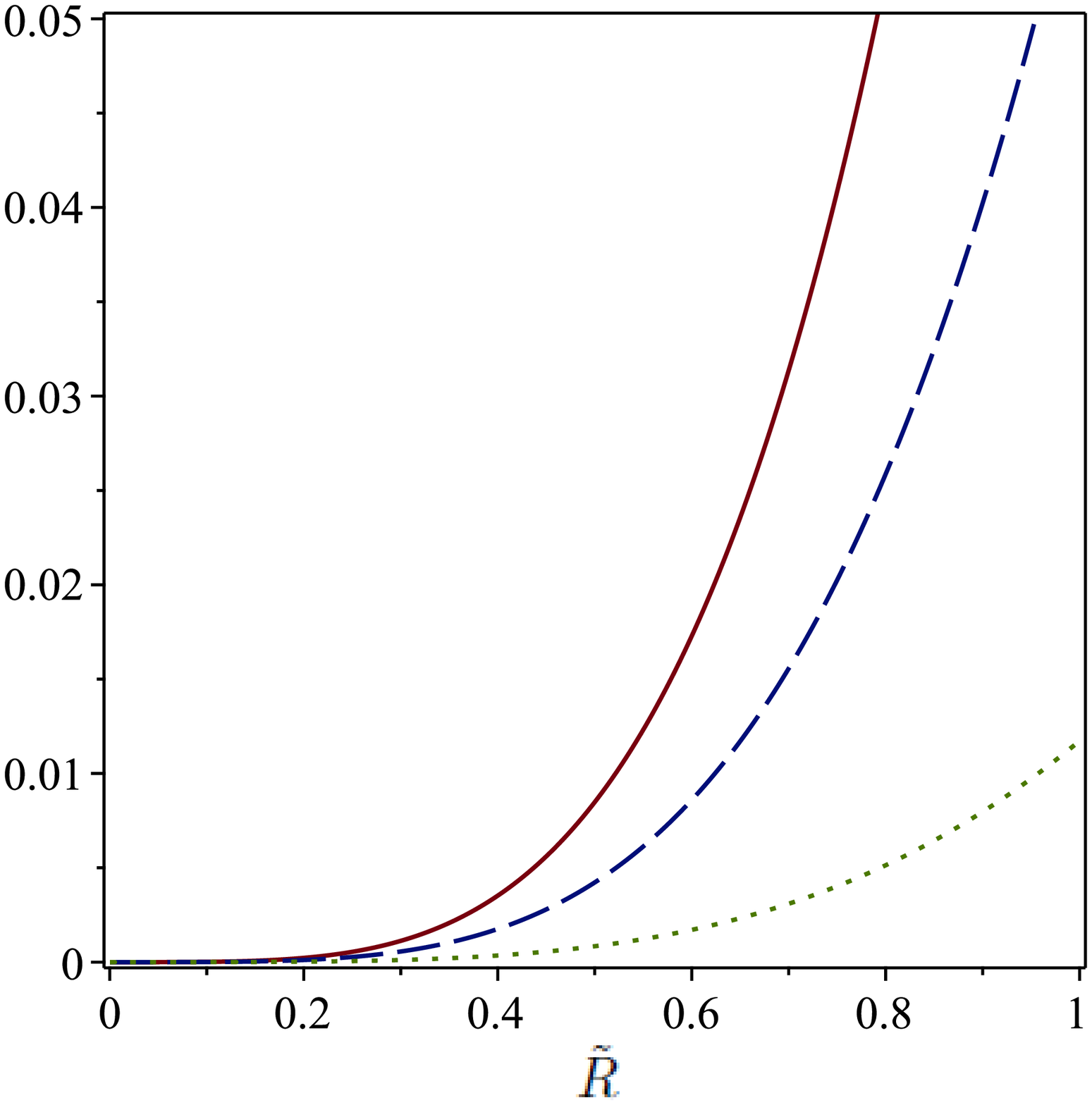}  \\  
 &  \\
(c)    &  (d)
\end{array}
$$	
\caption{The
specific angular momentum   $h^2_+$ and  $h^2_-$ for  charged test particles
around  Morgan-Morgan-like charged  dust thick disks with parameters  $\tilde e = 0$  (solid curves),  $0.5$,  $0.9$ (dotted curves),   $ \tilde d = 1$,   $\tilde b =  1$ (top figures) and   $\tilde b =  2$,   as functions of 
$\tilde R$ } 
\label{fig:fig6}
\end{figure}

In figure  \ref{fig:fig5} we graph the 
circular speed for direct orbits $v^2_+$ and  retrograde orbits  $v^2_-$  for  charged test particles moving 
around  Morgan-Morgan-like charged  dust thick disks with parameters  $\tilde e = 0$ (neutral particles),  $0.5$,  $0.9$ and the same values of the parameter $\tilde d$ and $\tilde b$,   as functions of  $\tilde R$.
We see that they  exhibit  the same behavior as the previous models.
However,  we observer that in these models  the particles are more relativistic   in the case  of  direct orbits,   whereas  for retrograde orbit the opposite occurs.   
Note also   that  circular speed  of particles is always a quantity
less than the speed  of light in agreement  with the dominant energy condition.  The orbits of particles are also stable against radial perturbations for these values of parameters  (figure   \ref{fig:fig6}).

\section{Morgan-Morgan-like charged dust  thick rings }
Ring like matter distributions  can be obtained  using a
Kelvin transformation \cite{Kelvin1,Kelvin2} which in cylindrical coordinates reads
\begin{equation}
(R, z) \rightarrow (\frac{d^2R}{R^2 + z^2},  \frac{d^2 z}{R^2 + z^2}). 
\end{equation}

When this transformation is applied to the relativistic  potential-density pair ($\psi$, $\rho$) corresponding to  the  non-linear  Poisson's equation (\ref{eq: rho}) we obtain  the  new pair 
\begin{subequations}\begin{eqnarray}
\bar \psi   & = & \left( \frac{d}{\sqrt{R^2 + z^2}} \right ) \psi \left( \frac{d^2 R}{R^2 + z^2},  \frac{d^2 z}{R^2 + z^2} \right) ,     \\
  \bar \rho   & = & \left( \frac{d^5}{(R^2 + z^2)^{5/2}} \right) \rho \left( \frac{d^2 R}{R^2 + z^2},  \frac{d^2 z}{R^2 + z^2} \right) .
\end{eqnarray}\end{subequations}
In particular, for the  potential-density pair (\ref{eq: rhoMM}) - (\ref{eq: psiMM}) we  obtain 
\begin{subequations}\begin{eqnarray}
\tilde \rho &= & \frac { 3  \tilde b^2 }{ 4  \pi \tilde d^3
  \tilde r ^5  (\bar x^2 + \bar y^2) \bar x^3   ( 1-\bar \psi )^3 },  
  \\
  \bar \psi  & =& -   \frac{1}{\tilde d \tilde r } \left \{ \cot^{ - 1} (\bar x) + \frac{1}{4}
 \left[ (3 \bar x^2  + 1) \cot^{ - 1} (\bar x) - 3 \bar x \right] \left( 3\bar y^2  - 1 \right) \right \} ,
\end{eqnarray}\end{subequations}
where  again $\tilde \rho =  M^2 \bar \rho $, $\tilde d = d/ M$,  $\tilde b = b/ d$,
\begin{subequations}\begin{eqnarray}
 \sqrt 2 \tilde r  \bar x  & = &  \sqrt { [ ( 1 + (\tilde b^2-1)\tilde r^2  )^2 + 4(\tilde z ^2 + \tilde b^2 \tilde r ^4)]^{1/2} 
 +  1 + (\tilde b^2-1)\tilde r^2} ,  \\
 \sqrt 2 \tilde r  \bar y  & = &  \sqrt { [ ( 1 + (\tilde b^2-1)\tilde r^2  )^2 + 4(\tilde z ^2 + \tilde b^2 \tilde r ^4)]^{1/2} 
 -  1 + (1-\tilde b^2)\tilde r^2}.
\end{eqnarray}\end{subequations}
and   $ \tilde r = \sqrt{ \tilde R^2 +  \tilde  z^2}$, being $\tilde R = R/d$ and $\tilde z = z/d$. 

\begin{figure}
$$
\begin{array}{cc}
\includegraphics[width=0.33\textwidth]{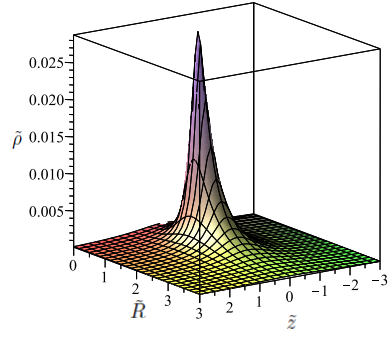} &   
\includegraphics[width=0.25\textwidth]{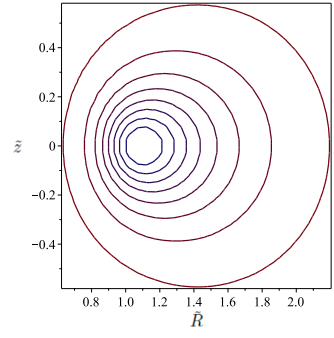}  \\  
 &  \\
(a)    &  (b)  \\
\includegraphics[width=0.33\textwidth]{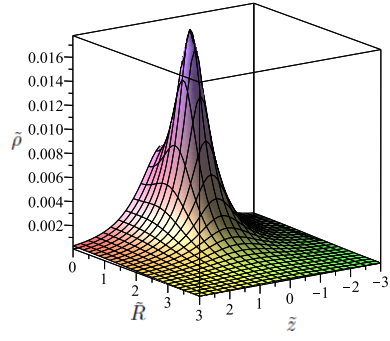} &   
\includegraphics[width=0.25\textwidth]{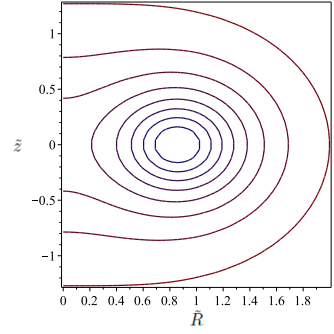}  \\  
 &  \\
(c)    &  (d)
\end{array}
$$	
\caption{The surface  and level curves of  the relativistic energy density $\tilde \rho$ for Morgan-Morgan-like charged dust thick rings  
with parameters
 $\tilde d = 1$, $\tilde b =  0.2$ (top figures) and  $\tilde b =  0.5$,   as functions of  $\tilde R$ and $\tilde z$. } 
\label{fig:fig7}
\end{figure}

\begin{figure}
$$
\begin{array}{cc}
v_+^2   &  v_-^2  \\  
\includegraphics[width=0.3\textwidth]{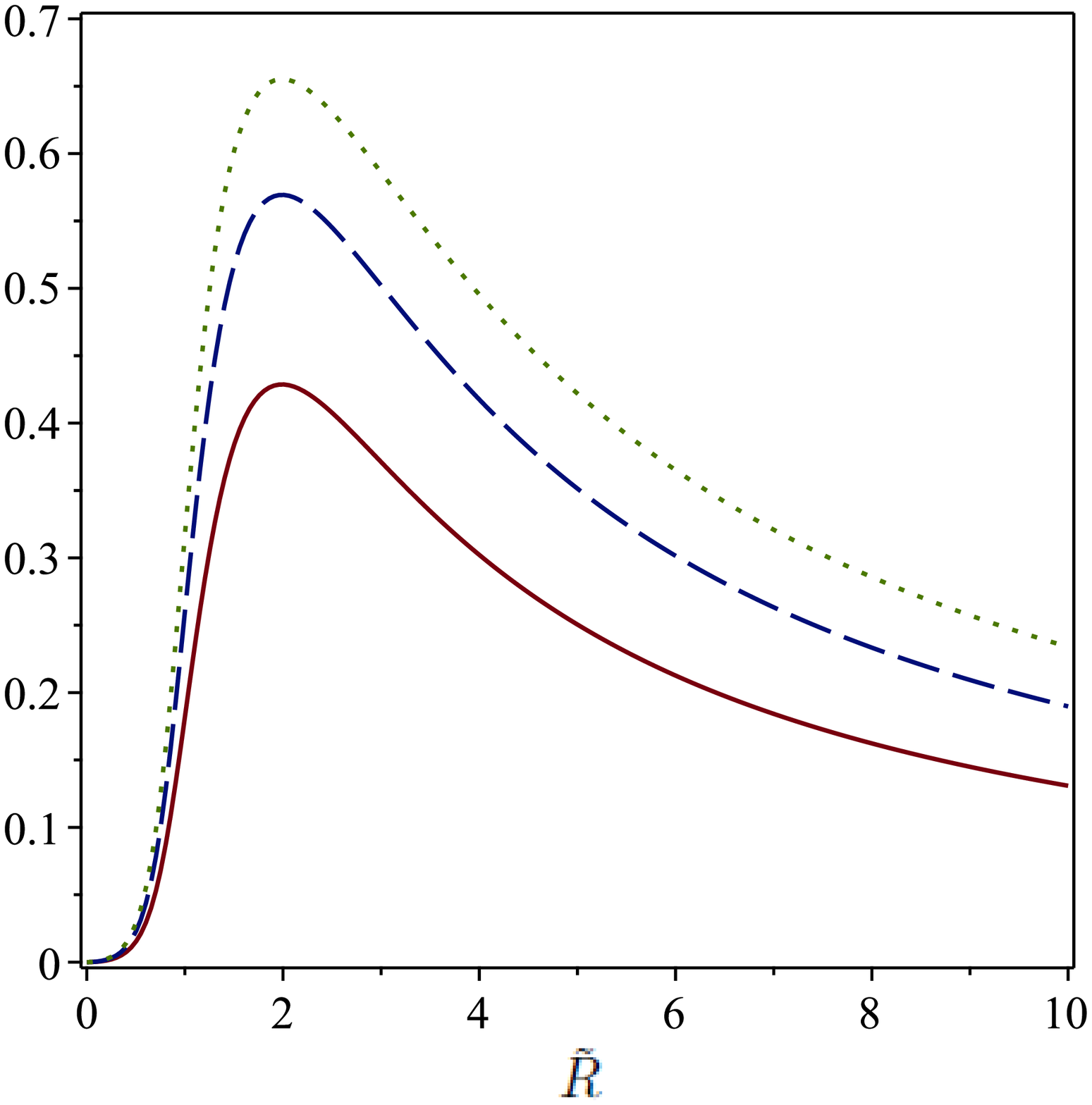} &   
\includegraphics[width=0.3\textwidth]{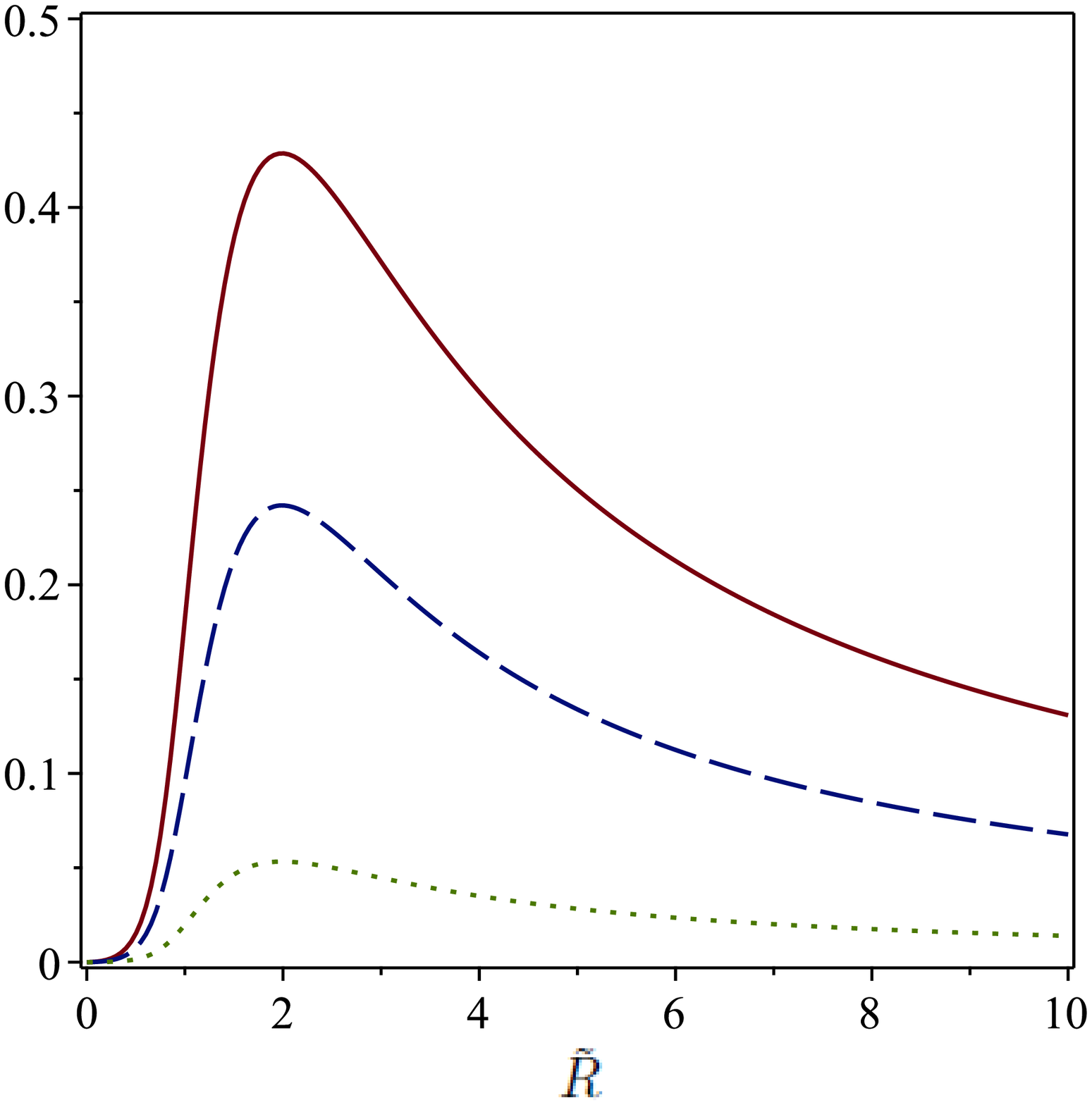}  \\  
 &  \\
(a)    &  (b)  \\
& \\
h_+^2   &  h_-^2  \\  
\includegraphics[width=0.3\textwidth]{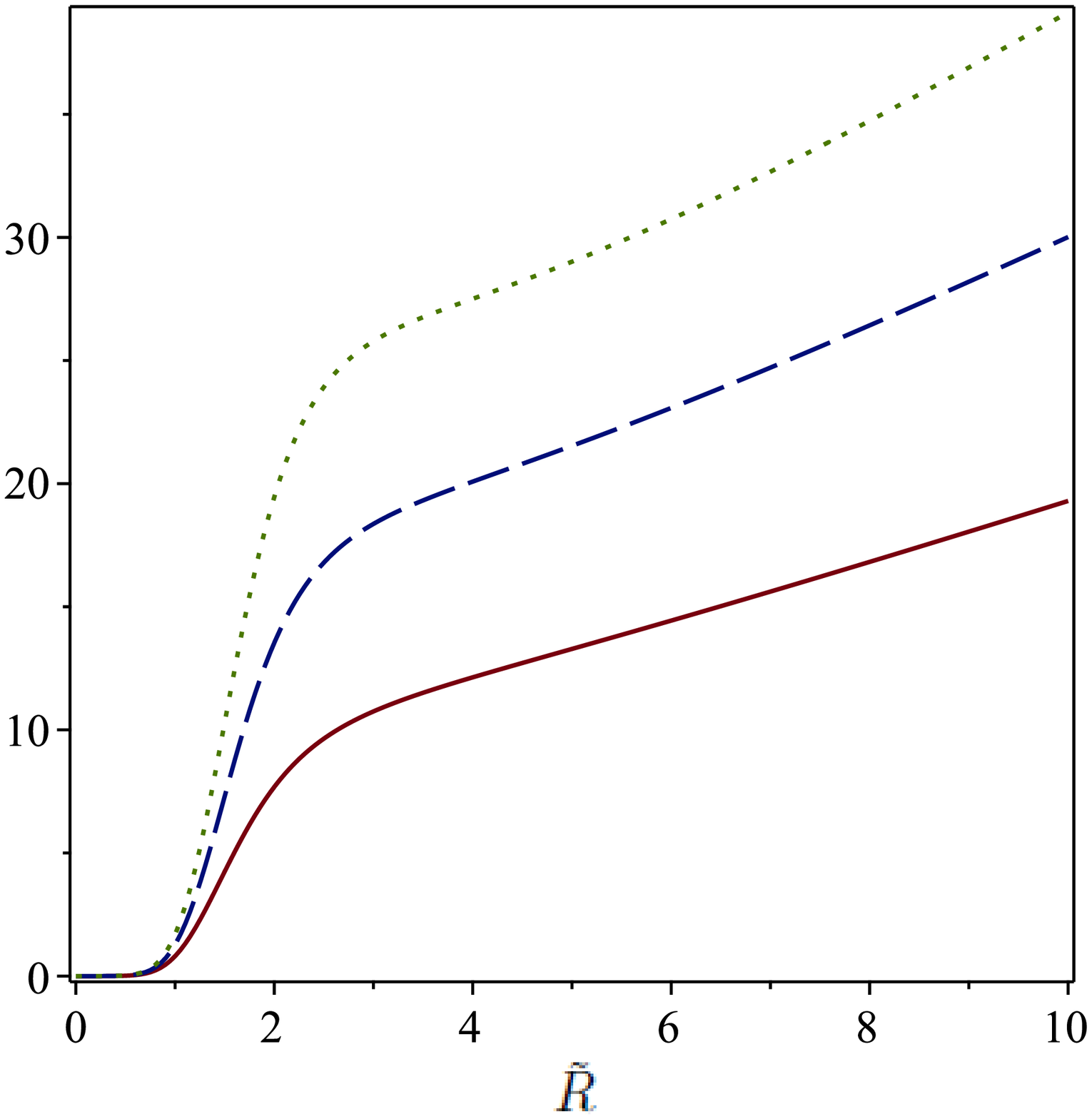} &   
\includegraphics[width=0.3\textwidth]{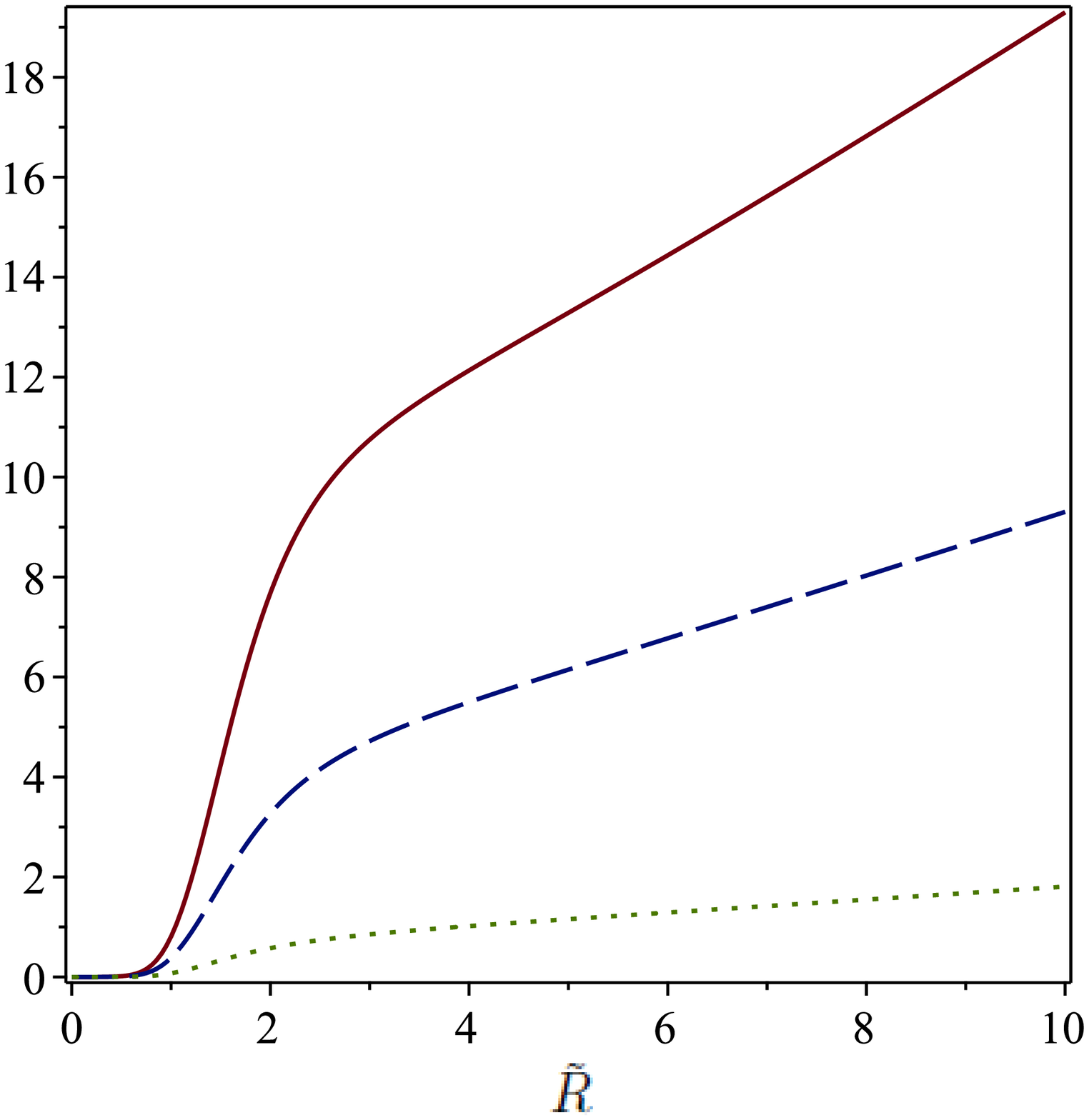}  \\  
 &  \\
(c)    &  (d)
\end{array}
$$	
\caption{ The
circular speed  $v^2_+$ and  $v^2_-$ (top curves) and the specific angular momentum   $h^2_+$ and  $h^2_-$  for  charged test particles
around   Morgan-Morgan-like charged dust thick rings with parameters  $\tilde e = 0$  (solid curves),  $0.5$,  $0.9$ (dotted curves),  $\tilde d = 1$,   $\tilde b =  0.5$,   as functions of 
$\tilde R$   } 
\label{fig:fig8}
\end{figure}

In figure \ref{fig:fig7} we show,  as functions of  $\tilde R$ and $\tilde z$, 
the  density profile  $\tilde \rho$ and contour plots for Morgan-Morgan-like charged dust  thick rings with 
 parameters  $ \tilde d = 1$, $\tilde b =  0.2$ and $\tilde b =  0.5$.   
The graphs suggest that we have a
ring structure.  In all cases we find that the energy density is a positive quantity.   In figure  \ref{fig:fig8}  we present the tangential speeds   $v^2_+$  and  $v^2_-$  for charged  test particles orbiting on the equatorial plane  and their   respective  specific angular momentums 
 $h^2_+$ and   $h^2_-$ for  parameters  $ \tilde d = 1$ and $\tilde b =  0.5$.  These quantities present the same behavior that in the previous models. However, for lower values of
 $\tilde b$, for example $\tilde b = 0.2$, one finds a central region
 where the  particles present  superluminal or tachyonic speeds and the orbits are unstable.  

\section{Discussion}

Two simple families of charged dust thick disks and a family of thick annular disks  for a particular conformastatic metric   were presented  based on the  extreme Reissner-Nordstrom black hole  field  and a Morgan-Morgan-like metric.
The disk models are constructed via a Miyamoto-Nagai transformation
used in Newtonian gravity to model flat galaxies and  the ring like matter distributions   applying  a Kelvin transformation.  The structures  satisfy all the energy conditions. Unlike  the  Morgan-Morgan-like  thin disks, the  relativistic thick disks built from them    have infinite extension, but as  the energy density   decreases rapidly with radius $\tilde R$   we can  to model such  structures  as  compact objects. 

We analyzed the electrogeodesic equatorial circular motion of charged test particles around of the disks.
We found that for direct orbits  the specific electric charge  increases the speed of the particles whereas
for retrograde orbits the contrary occurs.      Also,  for direct orbit the particles  become more relativistic  when the matter distribution is flatted, whereas
for retrograde orbits   the opposite  occurs.  In the  Morgan-Morgan-like disk models the speed of  the particles is greater  in the case  of  direct orbits,     whereas  for retrograde orbit the opposite occurs.
In all the cases,  we found stable orbits  against radial perturbations.


\newpage


\section*{References}

\begin{thebibliography}{999} 


\bibitem{BS} W. A. Bonnor and A. Sackfield, Commun. Math.   Phys. {\bf 8}, 338
(1968).

\bibitem{MM1} T. Morgan and L. Morgan, Phys. Rev.  {\bf 183},  1097 (1969).

\bibitem{MM2} L. Morgan and T. Morgan, Phys. Rev. D  {\bf 2},  2756 (1970).



\bibitem{LP} D. Lynden-Bell and S. Pineault, Mon. Not. R.  Astron. Soc. {\bf
185}, 679 (1978). \label{bib:LP}

\bibitem{CHGS} A. Chamorro, R. Gregory, and J. M. Stewart, Proc. R. Soc. London
{\bf A413}, 251 (1987).

\bibitem{LO} P.S. Letelier and S. R. Oliveira, J. Math.  Phys.  {\bf 28}, 165
(1987).

\bibitem{LEM} J. P. S. Lemos, Class. Quantum Grav. {\bf 6}, 1219 (1989).

\bibitem{BLK} J. Bi\u{c}\'{a}k, D. Lynden-Bell, and J.  Katz,  Phys. Rev. D {\bf
47}, 4334 (1993).

\bibitem{BLP} J. Bi\u{c}\'{a}k, D. Lynden-Bell, and C.  Pichon, Mon. Not. R.
Astron. Soc. {\bf 265}, 126 (1993).

\bibitem{GE} G. A. Gonz\'alez and O. A. Espitia,  Phys. Rev. D  {\bf 68}, 104028
(2003).



\bibitem{BL} J. Bi\u{c}\'ak and T. Ledvinka, Phys. Rev.  Lett. {\bf 71}, 1669
(1993).

\bibitem{GL2} G. A. Gonz\'alez and P. S. Letelier, Phys.  Rev.  D {\bf 62},
064025 (2000).


\bibitem{LL1} J. P. S. Lemos and P. S. Letelier, Class.  Quantum Grav. {\bf
10}, L75 (1993).

\bibitem{LL2} J. P. S. Lemos and P. S. Letelier, Phys. Rev. D  {\bf 49},  5135
(1994).

\bibitem{G-L-thick}  G. A. Gonz\'alez and P. S. Letelier, Phys.  Rev.  D {\bf 69}, 044013 (2004).

\bibitem{V-L-thick}  D. Vogt and   P. S. Letelier,  Phys. Rev. D  {\bf 76}, 084010   (2007).



\bibitem{LBZ} T. Ledvinka, J. Bi\u{c}\'{a}k, and M.  \u{Z}ofka, in {\it
Proceeding of 8th Marcel-Grossmann  Meeting in General Relativity}, edited by
T. Piran  (World  Scientific, Singapore, 1999)

\bibitem{KBL} J. Katz, J. Bi\u{c}\'ak, and D. Lynden-Bell, Class. Quantum Grav.
{\bf 16}, 4023 (1999).


\bibitem{LET1} P. S. Letelier, Phys. Rev. D {\bf 60},  104042  (1999).

\bibitem{GG1} G. Garc\'\i a R. and G. A. Gonz\'alez, Phys.  Rev. D  {\bf 69},
124002 (2004).

\bibitem{GG2} G. Garc\'\i a-Reyes and G. A. Gonz\'alez,  Class. Quantum Grav.
{\bf 21}, 4845 (2004).

\bibitem{GG-CRIS} C. H. Garc\'{\i}a-Duque and G.  Garc\'{\i}a-Reyes, Gen.
Relativ. Gravit. {\bf 43}, 11, 3001 (2011).

\bibitem{G-O} G. Garc\'\i a-Reyes and O. A. Espitia,  Gen. Relativ. Gravit. {\bf 46}, 1674 (2014).


\bibitem{Rosseland} S. Rosseland, Mon. Not. R. Astron. Soc. {\bf 84}, 720 (1924).

 \bibitem{Bally} J. Bally and E. R. Harrison, Astrophys. J.   {\bf 220}, 743 (1978).

\bibitem{V-L-perfect} D. Vogt and   P. S. Letelier,  Phys. Rev. D  {\bf 70}, 064003   (2004).


\bibitem{majum} S. D. Majumdar,  Phys. Rev. {\bf 72},  390 (1947).

\bibitem{papa} A. Papapetrou, Proc. Roy. Soc. (London)  {\bf A51}, 191 (1947).

\bibitem{bonnor98} W. A. Bonnor,  Class. Quantum Grav. {\bf 15}, 351 (1998). 

\bibitem{V-L-dust} D. Vogt and   P. S. Letelier,  Class. Quantum Grav. {\bf 21},  3369  (2004). 

\bibitem{G-A-P} G. A. Gonz\'alez, A. C. Guti\'errez-pi\~neres and P. A. Ospina,  Phys. Rev. D  {\bf 78}, 064058   (2008).

\bibitem{Lora} F. D. Lora-Clavijo, P. A. Ospina-Henao and J. F. Pedraza, Phys. Rev. D {\bf 82}, 084005  (2010).   


\bibitem{Miyamoto}  M.  Miyamoto and   R. Nagai, PASJ  {\bf 27}, 533 (1975).

\bibitem{Nagai} R. Nagai and M.  Miyamoto,  PASJ {\bf 28}, 1 (1976).

\bibitem{Kuzmin} G. G. Kuzmin 1956,  Astron. Zh.  {\bf 33}, 27 (1956). 

\bibitem{Toomre} A. Toomre, Ap. J.  {\bf 138}, 385  (1962). 


\bibitem{Kelvin1} W. Thomson (Lord Kelvin), J. Math. Pures Appliquees {\bf 12}, 256 (1847).

\bibitem{Kelvin2}  O. D.  Kellog, {\it  Foundations of Potential Theory},  Dover Publications, New York (1953). 

\end{thebibliography}
\end{document}